# Dynamical System Modeling to Simulate Donor T Cell Response to Whole Exome Sequencing-Derived Recipient Peptides Demonstrates Different Alloreactivity Potential In HLA-Matched and Mismatched Donor-Recipient Pairs

Abdul Razzaq B; Scalora A; Koparde VN; Meier J; Mahmood M; Salman S; Jameson-Lee M; Serrano M; Sheth N; Voelkner M; Kobulnicky DJ; Roberts CH; Ferreira-Gonzalez A; Manjili MH; Buck GA; Neale MC; Toor AA.

Bone Marrow Transplant Program, Department of Internal Medicine; Center for Biological Complexity; Department of Psychiatry and Statistical Genomics Virginia Commonwealth University.

Correspondence: Amir A. Toor MD, Associate Professor of Medicine, Bone Marrow Transplant Program, Massey Cancer Center, Virginia Commonwealth University, Richmond, VA 23298. Ph; 804-628-2389. E-mail: amir.toor@vcuhealth.org

Key Words: Stem cell transplantation, Whole exome sequencing, GVHD, Dynamical system, HLA, Logistic Equation, Vectors, Operators, Histocompatibility

**Abstract.**

Immune reconstitution kinetics and subsequent clinical outcomes in HLA matched recipients of allogeneic stem cell transplantation (SCT) are variable and difficult to predict. Considering SCT as a dynamical system, may allow sequence differences across the exomes of the transplant donors and recipients to be used to *simulate* an alloreactive T cell response, which may allow better clinical outcome prediction. To accomplish this, whole exome sequencing was performed on 34 HLA matched SCT donor-recipient pairs (DRP), and the nucleotide sequence differences translated to peptides. The binding affinity of the peptides to the relevant HLA in each DRP was determined. The resulting array of peptide-HLA binding affinity values in each patient was considered as an *operator* modifying a hypothetical T cell repertoire *vector*, in which each T cell clone proliferates in accordance to the logistic equation of growth. Using an iterating system of matrices, each simulated T cell clone's growth was calculated with the steady state population being proportional to the magnitude of the binding affinity of the driving HLA-peptide complex. Incorporating competition between T cell clones responding to different HLA-peptide complexes reproduces a number of features of clinically observed T cell clonal repertoire in the simulated repertoire. These include, sigmoidal growth kinetics of individual T cell clones and overall repertoire, Power Law clonal frequency distribution, increase in repertoire complexity over time with increasing clonal diversity and finally, alteration of clonal dominance when a different antigen array is encountered, such as in stem cell transplantation. The simulated, alloreactive T cell repertoire was markedly different in HLA matched DRP. The patterns were differentiated by rate of growth, and steady state magnitude of the simulated T cell repertoire and demonstrate a possible correlation with survival. In conclusion, exome wide sequence differences in DRP may allow simulation of donor alloreactive T cell response to recipient antigens and may provide a quantitative basis for refining donor selection and titration of immunosuppression following SCT.

**Introduction.**

Stem cell transplantation (SCT) from HLA matched donors delivers curative therapy to patients with hematological malignancies, however at the cost of significant morbidity derived from the alloreactive phenomenon of graft versus host disease (GVHD). A major limitation in optimizing clinical outcomes following SCT has been the inability to control the onset and severity of GVHD without intensive immunosuppression. The more immunosuppression is increased in the peri-transplant period the more likely it becomes that GVHD risk will be diminished, but reciprocally the risk of complications such as opportunistic infections and disease relapse goes up, eventually neutralizing the survival benefit from lower GVHD risk. [1, 2, 3] On the other hand, rapidity and completeness of immune reconstitution helps to mitigate these complications but may be associated with a higher risk of GVHD in some individuals. [4, 5] Because of the complexity of the variables involved the occurrence of GVHD has been dealt with as a probability function of various clinical features of the transplant. Donor type, histocompatibility, transplant donor recipient gender matching, minor histocompatibility antigens (mHA), intensity of immunosuppression before and after transplantation have figured prominently in such analyses. [6, 7, 8] HLA matching of donors and recipients, mHA differences and MHC locus similarity have stood out as the most critical determinants of GVHD onset and clinical outcomes in such analyses. However, within the constraints of these parameters it is still difficult to precisely predict the development of alloreactivity manifesting as GVHD in individuals undergoing SCT. [9]

Precise control cannot be exercised without knowledge of the entire library of antigens driving post-transplant immune reconstitution. This task is made difficult by the multi-organ nature of GVHD, which makes it likely that the underlying mHA target landscape in HLA matched pairs is varied and complex, and the ensuing T cell response equally as diverse. Further, relatively simple changes in immunosuppressive regimens (to neutralize HLA-mismatch-directed alloreactivity), have allowed transplantation to be performed using donors with increasing levels of disparity across the HLA loci; best exemplified by related, HLA-haploidentical SCT and unrelated, umbilical cord blood transplantation (UCBT). [10] In these examples of HLA mismatched transplants, GVHD rates are similar or lower when compared with HLA identical donors. While these differences are partly a function of the immunosuppressive regimens used and partly the lack of a mature T cell repertoire, they point to the possibility that there does *not* exist a quantitative relationship between the extent of HLA matching in transplant recipients and donors, and clinical outcomes. Further, the apparently random nature of occurrence of alloreactivity in similarly HLA matched (or mismatched) individuals, points to the influence of genetic disparity across the entire exome (as the determinant of mHA) on clinical outcomes.

*Histocompatibility Antigens In SCT*

Classically matching for the *major histocompatibility* (MHC) loci has been the cornerstone of stem cell donor selection. From an immunological standpoint, *alloreactivity* in recipients of HLA matched SCT is driven in part by histo-incompatibilities between transplants donors and recipients, differences in mHA. In patients matched for the MHC loci, these mHA generally provide the trigger for initiation of alloreactive tissue injury by donor T cells. Previous work has painstakingly identified numerous HLA specific mHA in transplant donor-recipient pairs (DRP). Drawing upon this background, a simple whole exome sequencing (WES) approach has demonstrated large numbers of potentially immunogenic DNA sequence differences between transplant donors and recipients. [11] These DNA sequence differences when translated to amino-acid sequences yield putative peptides, which demonstrate a range of binding affinities for the HLA molecules present in a specific donor and recipient. This WES + *in silico* HLA-peptide binding affinity procedure yields a large DRP-specific library of putative mHA which may contribute to alloreactivity resulting in GVHD (HLA binding peptides absent in donors but present in recipients). [12] Compared to the experimentally-determined mHA, this computational procedure yields several hundred to thousands of *potential mHA* in each HLA-matched DRP, and thus introduces a very large set of variables which cumulatively may influence the likelihood of GVHD incidence in individual SCT recipients. Factors contributing to the influence of these putative mHA on GVHD incidence will include tissue localization and levels of expression of the proteins that these peptides are derived from. Aside from such *antigen abundance* parameters, protein glycosylation and the presence of peptide cleavage sites will also influence the *immunogenicity* of the putative mHA in specific DRP. So while this *derived* variant HLA binding peptide library does not reflect the *entire* histocompatibility antigen spectrum, it gives an *estimate* of the scale of *antigen diversity* encountered by the donor T cells when transplanted into the recipient, and may be considered as a *estimated alloreactivity potential*. This measure of antigenic diversity may then be studied to lend an insight into the immunobiology of transplants. However, the uniformly large magnitude of this measure of alloreactivity potential precludes it being very useful in donor selection beyond the current standard of HLA matching. To do so the influence of these putative mHA on immune reconstitution must be understood, in other words, the T cell response to the mHA library in each patient must be determined.

The large array of potential peptide targets of alloreactivity, each with a different binding affinity to HLA, is presented on a limited number of HLA class I (present endogenous antigens), and HLA class II (present exogenous antigens) molecules in the recipient tissues. This makes it very unlikely that GVHD incidence is a probability function of a small number of critical variables, rather it is more likely that GVHD is the result of an integrated sum of all the variant immunogenic peptides presented by the HLA in a unique DRP, modulated by the immunosuppressive influence of pharmacotherapy during immune reconstitution. In other words the putative mHA serve as targets of GVHD and HLA molecules as modulators of this condition; GVHD then results as a consequence of T cell (and B cell) responses to these variant antigen targets. This is a logical assumption since donor T cell reconstitution corresponds with the likelihood of alloreactivity developing. [4, 13, 14]

In order to develop a model to estimate the likelihood of alloreactivity developing in unique SCT-DRP, immune response to the peptide-HLA complex library will need to be modeled in each individual. In such a model the binding affinity of the mHA to the relevant HLA molecules will be one of the determinants of the likelihood of individual mHA being presented to T cells. Further, unlike the conventional wisdom of the receptor-ligand relationship the peptide-HLA binding affinity distribution observed in this computationally-derived dataset is not discrete, rather it is continuous and non-linear, which suggests that T cell response may be similarly scaled.

*T Cell Repertoire Following SCT*

T cell repertoire following SCT is comprised of a complex collection of T cell clones, each of which expresses unique T cell receptors (TCR), predominantly dimers of TCR $\alpha$ and $\beta$, which in turn display a varying degree of avidity for peptide-HLA complexes. Like the peptide-HLA complexes observed there are thousands of T cell clones which span a large range of clonal frequencies, with a slightly different range of frequencies for different T cell clones observed in each individual. It is important to note that the TCR are generated by rearrangement of a set of variable, (diversity) and joining segments in the genomic loci for the TCR. These rearranged loci yield receptor molecules which heterodimerize to generate T cell clones, which over a lifetime are selected to yield a self-tolerant, pathogen-responsive set of clones which responds to pathogen-derived peptides in the context of an individual's HLA molecules. Clones, which encounter antigen are either propagated or suppressed based on whether or not they are autoreactive. The T cell receptors are coupled with CD3 molecules, which deliver a proliferation/activation signal to the T cells when the *first signal* target antigen-HLA complex is encountered. Therefore, depending on the presence and immunogenicity of the target antigens, T cell clones may be abundant or scarce. An important moderator of the immune responsiveness is the presence of the *second signal*, which either enhances or extinguishes immune responsiveness (CD28 and PD-1 respectively). These in turn influence the response observed to the antigens the T cell may recognize. Within T cells there are several functionally unique subsets, with cytotoxic, helper and regulatory function. These are characterized by expression of molecules such as CD8 or CD4 co-receptors, and IL-2 receptor $\alpha$ which in turn ligate the non-peptide binding domains of HLA class I or class II molecules, or serve as binding sites for cytokines. Each of these functional subsets possesses a unique array of TCR $\alpha\beta$, and can recognize and respond to a unique array of HLA-peptide complexes. The eventual T cell repertoire is massive, potentially numbering in the several hundred thousand to more than a million unique T cell clones. [15, 16, 17]

High throughput sequencing of TCR cDNA obtained from isolated T cells has made it possible to determine the clonal frequency of distinct T cell populations and demonstrates, continuous, non-linear distribution. In this instance the distribution of T cell clonal frequency when ordered according to clonal rank conforms to a Power Law distribution, declining as a power of the number of clones examined. [18, 19, 20, 21] These clonal frequencies may change under different conditions, such as upon encountering pathogens in normal circumstances or in SCT upon encountering alloreactive-antigens. Specifically following SCT the T cell clonal repertoire in the donor product is altered, with clones that were once dominant becoming suppressed and other previously repressed clones becoming ascendant. [18, 22]

The concordance between the peptide-HLA binding affinity distribution and the T cell clonal frequency distribution may be further analyzed quantitatively. Understanding the quantitative principles at hand will lend new insights into the variable presentation of GVHD in HLA matched and mismatched individuals, both from the standpoint of incidence as well as the extent and severity of involvement. Lymphocyte reconstitution following SCT may be modeled as a *logistic dynamical system*, with a familiar sigmoid growth pattern over time. [4, 23] Logically then, various T cell subsets, including T cell clones should demonstrate the same growth dynamic. This would be consistent with the notion of self-similarity of T cell clonal distribution within the T cell repertoire of an individual. This logical assumption makes it possible to develop a model, which may predict immune reconstitution following SCT, by simulating the T cell repertoire emerging in response to the mHA-HLA complex library derived from DRP exome differences.

*Dynamical System Modeling Of T Cell Response To Putative mHA*

The ability to model donor T cell repertoire response to the magnitude of recipient antigen diversity will be an important step towards predicting the GVHD causing potential of potential transplant donors. The first step in this process is to understand the basic rule that governs T cell response to antigens, i.e., proliferation and growth. Postulating that immune effectors constitute a system that follows mathematically defined laws, it may be modeled as a *dynamical system* [24, 25, 26, 27]. A dynamical system is an *iterating* physical system, where the variable being studied evolves over time in a manner such that state of the variable in the system at any given time, is predicated on all the preceding states, and the evolution of the system can be modeled using *ordinary differential equations*, which for systems demonstrating logistic growth will take the general form

$$\frac{dN}{dt} = rN_t \left( \frac{K - N_t}{K} \right)$$

Here, *dN/dt* is the instantaneous rate of change of the cell population at time *t*, when the population $N_t$ has an intrinsic growth rate, *r* and a maximum growth potential to reach an eventual size, *K*. This logistic dynamical system constitutes a system of repeated calculations, where the output of the equation for each iteration gives the population $N_t$, as a *function* of time, and becomes the input variable (*argument*) for the next calculation $N_{t+1}$. This system behaves in a *non-linear* fashion, demonstrating sigmoid population growth constrained by feedback and by environmental pressures (Supplementary Figure 1). A feature of logistic dynamical systems is that the starting parameters for the system such as the initial population size, and growth potential, set the trajectory of the series of events to follow, and thus determine the eventual outcome. Small variations in the early state of the system can produce large measurable effects late in the evolution of this system.

In this paper we present a model of T cell reconstitution based upon the application of the logistic dynamical system to a hypothetical T cell repertoire responding to donor recipient alloreactivity potential as previously described. [12] For this study exome sequencing was performed on DNA from 34 SCT donor-recipient pairs, with the resulting data computationally translated to yield putative variant peptides with a graft versus host (GVH) direction (variant amino acid present in the recipient but absent in the donor). HLA class I binding affinity of nona-meric peptides to the relevant HLA in the DRP was determined. The resulting data were used in a mathematical model of CD8+ T cell reconstitution, where T cell clonal proliferation would be proportional to the antigen binding affinity. This model demonstrates significant differences in the *alloreactive T cell repertoire* evolving in different HLA matched DRP, suggesting that genetic back ground may exert a substantial impact on post transplant T cell reconstitution, possibly explaining disparate outcomes seen in equivalently HLA matched SCT.

**Methods.**

*Logistic Dynamical System*

If SCT is considered as a system of interacting donor T cell clones and recipient antigens, immune reconstitution following SCT may be considered as a dynamic, evolving process, which can be modeled mathematically. This may allow more precise determination of the odds of clinical outcomes such as engraftment, GVHD and survival in an individual [23]. In this model, each T cell clone responding to its cognate antigen will proliferate, conforming to the *iterating logistic equation* of the general form

$$N_t = \frac{K}{(K - N_{t-1})(e^{rt}) + 1} \qquad [1]$$

In this equation, $N_t$, is the T cell count at time $t$ following transplant (modeled as iterations); $N_{t-1}$ represents the T cell count in the previous iteration; $K$ is the T cell count at steady state reached after several iterations, in other words the maximum T cell count the system would support (carrying capacity); $r$, is the growth rate of the population and $t$ is the number of iterations, that the population has gone through (representing time). As noted above, the T cell population, $N_t$, at time, $t$ following SCT, is a function of the T cell population $N_{t-1}$, at an earlier time point, $t-1$. This in turn will be dependent on the T cell population at the outset of transplant $N_0$, and the intrinsic T cell proliferative capacity (growth rate) $r$ governing the growth, such that

$$N_0 \xrightarrow{[r]} N_{t-1} \xrightarrow{[r]} N_t \dashrightarrow^{[r]} K$$

In this model, the final steady state population of all the T cell clones ($n$) ($\sum_1^n K$), and its clonal repertoire will determine clinical outcomes. This final repertoire will be largely dependent on the antigen responsiveness of the clones, which would determine the $K$ for each clone. Other formative influence that may be added to the model will be antigen abundance and ambient cytokine milieu + pharmacotherapy at the time of T cell growth.

*Matrices To Model The T Cell Clonal Repertoire: Vector Spaces And Operators*

While it is easy to apply the logistic equation to describe the evolution of the whole system, *as an average*, applying this concept to all the individual T cell clones, in order to simulate the evolution of the T cell clonal repertoire from the putative mHA library for a specific DRP, is not straightforward. Such an analysis would allow the capability of *simulating* alloreactive T cell response that one donor might have as opposed to a different one, based on the exome variation between them and the recipient in question, with the relevant HLA background. The mHA-HLA binding affinity distribution may be utilized to do so, if a number of simplifying assumptions are made.

The T cell clones present in an individual can be considered as a set of individual *vectors* in the immune *phase space* of that individual. The description vector in this instance may also be expanded to describe the entire set of T cells, with the individual T cell clones representing components of this vector. [28] In this vector, T cell clonality represents direction (since it determines antigen specificity) and T cell clonal frequency, the magnitude of individual vector components. The sum of these vectors will then represent the average T cell vector and its overall direction, either tolerance or alloreactivity. The T cell vectors may take a number of different configurations in an individual, and the entire range of vector configurations possible will constitute the immune phase space for that individual. The T cell vector may be represented mathematically as a single column matrix, with T cell clonal frequencies $TC_1$, $TC_2$, $TC_3$ …. $TC_n$. This T cell clonal matrix may be represented by the term, $\overline{v_{TCD}}$. The frequency of each TC is a natural number, and the direction, its reactivity, represented by the clonality. Phase space will then describe all the potential T cell clonal frequency combinations possible, within a specific antigenic background, with the all the potential directions of the T cell reactivity. The infused T cell clonal repertoire will represent a steady state T cell clonal frequency distribution in the normal donor, presumably made up of mostly self-tolerant, pathogen specific T cells, and without any auto-reactive T cells. This represents a unique state, or configuration of the T cell vector in the T cell phase space of that individual, which will be altered upon transplantation. For the computations reported here, each T cell clone has a $N_0$ value of 1, and is alloreactive.

Once the SCT occurs, the infused T cells encounter a new set of antigens (both recipient and pathogen). These antigens presented by the recipient (or donor-derived) antigen presenting cells constitute an *operator*, a matrix of targets that the donor T cells may proliferate in response to. [29] This operator changes the magnitude and direction (clonal dominance) of the infused T cell repertoire, as individual T cell clones in the donor product grow or shrink in the new HLA-antigen milieu, altering the T cell vector configuration, in other words transforming it. The putative mHA making up the alloreactivity potential constitute a matrix, which may be termed an *alloreactivity potential operator,* $\mathbb{M}_{APO}$. This operator will consist of the binding affinity of the unique variant peptide-HLA complexes likely to be encountered by donor T cells in an individual donor. Following SCT, the donor T cell clonal frequency shifts according to the specificity of the TCR and the abundance and reactivity of the corresponding antigen. This corresponds to the operator modifying the underlined *original* donor T cell vector, $\overline{v_{TCD}}$ as the system goes through successive iterations (Supplementary Figure 2), to the *new recipient* T cell vector, $\overline{v_{TCR}}$, according to the following relationship

$$\overline{v_{TCD}} \cdot \mathbb{M}_{APO} = \overline{v_{TCR}} \qquad [2]$$

*Applying the Logistic Growth Equations to Vector-Operator Systems*

A central assumption in this model is that the steady state TC clonal frequency ($K$) of specific TCR bearing clones, and their growth rates ($r$), are proportional to the binding affinity of their target mHA-HLA complexes. This is *approximately* represented by the reciprocal of the *IC50* (test peptide concentration in *nM* required to displace a standard peptide from the HLA molecule in question) for that specific complex. This is because the strength of the binding affinity will increase the likelihood of T cell-APC interactions occurring, as well as the driving the rate of this interaction. While each TCR may recognize another mHA-HLA complex with equal or lesser affinity, for the sake of simplicity, an assumption of non-recognition of other mHA-HLA complexes will be made for this model. This makes it possible to construct a square *identity matrix* with unique mHA-HLA complexes for a SCT-DRP and the corresponding TCR, where $TCR_x$ recognizes $mHA_x$-HLA (1) and not others (0). In addition to the assumption of there being a first signal of TCR recognition, a second signal immune responsiveness is assumed in this matrix.

| SCT | | $\mathbb{M}_{APO}$ | | | |
|---|---|---|---|---|---|
| | | $mHA_1$ HLA | $mHA_2$ HLA | $mHA_3$ HLA | $mHA_n$ HLA |
| $\overline{v_{TCD}}$ | $TC_1$ | 1 | 0 | 0 | 0 |
| $\downarrow t$ | $TC_2$ | 0 | 1 | 0 | 0 |
| $\overline{v_{TCR}}$ | $TC_3$ | 0 | 0 | 1 | 0 |
| | $TC_n$ | 0 | 0 | 0 | 1 |

The *alloreactivity matrix* modifies the donor T cell clonal vector infused with an allotransplant, mapping it to the recipient T cell vector, as the T cell clones with unique TCR encounter the corresponding mHA-HLA complexes they proliferate conforming to the logistic equation [1]. In the logistic equation $K$ for each T cell clone will be proportional to the approximate binding affinity ($1/IC50$) of the corresponding mHA-HLA complex ($K^{1/IC50}$). Further, $K$ will be a multiple of the relative peptide antigen abundance (expression levels in various tissues), so a highly expressed protein derived alloreactive peptide with a low affinity, in a large volume target organ may yet produce a correspondingly large T cell response, despite the low affinity for the TCR and *vice versa* (Supplementary Figure 3A). The $r$ will be a function of the binding affinity (by increasing the TCR-mHA-HLA interaction time) and the intrinsic T cell proliferation capacity. Equation 1 will therefore be modified for unique TC clone $TCx$, responding to *mHAx-HLA*, as follows,

$$N_{t\ TCx} = \frac{K^{1/IC50mHAx}}{(K^{1/IC50\ mHAx} - N_{t-1\ TCx})(e^{rt(1/IC50mHAx)}) + 1} \quad [3]$$

In real life, local effects such as tissue injury and resulting antigen presentation will also form critical contributors to alloreactivity manifesting as GVHD. The other major influence on the alloreactivity will be the ambient cytokine concentration and immunosuppressive pharmacotherapy. These are not modeled in the present work.

In this alloreactivity, vector-operator identity matrix, the values 1 and 0 in each cell will be multiplied by the product of the logistic equation, equation 3, for each T cell clone. The T cell response to each mHA-HLA complex is determined over time, $t$, by iterating the system of matrix-equations. In this *alloreactivity matrix*, the IC50 of all the alloreactive peptides with a GVH direction (present in recipient, but absent in donor) constitutes the operator; the sum of $n$ T cell clones $\sum_{1}^{n} TC$, at each time point will represent the magnitude of the vector $\overline{V_{TCR}}$ at that time $t$. In this system, when considering the effect on infused donor T cell vector, depending on antigen affinity, T cell clones present in abundance may be down regulated and other clones present at a low number may go up exponentially upon encountering antigen, *transforming* the vector over time from the original infused T cell vector.

It is to be noted that either tolerance induced by immunosuppression, or removal of putative T cell clones by T cell depletion will change a 1 to a 0 in the alloreactivity matrix, reducing the degree of transformation sustained by the donor T cell vector.

In summary, the alloreactivity operator determines the expansion of the donor T cell vectors, following transplantation, in an iterating fashion, *transforming* it to a new configuration, over time *t*, based on the mHA-HLA complexes encountered and their affinity distribution and antigen abundance in the recipient. Therefore post SCT immune reconstitution may be considered as a process in which T cell clonal frequency vectors are multiplied by the minor histocompatibility antigen matrix operator in an iterating fashion and this results in transformation of the vector over time to either a GVHD-prone alloreactive or to a tolerant, pathogen-specific vector. This may be visualized as thousands of T cell clones interacting with antigen presenting cells, an example of interacting dynamical systems previously reported. [4]

*Competition between T Cell Clonal Populations*

Each of the T cell clones is behaving like a unique population, therefore *competition* with other T cell clones in the set of all T cell clones must be accounted for in the logistic equation to determine the magnitude of the unique clonal frequencies as the model iterates simulating T cell clonal growth over time. This may be done using the *Lotka-Volterra* correction for *competing populations*, which accounts for the impact of population growth of multiple coexisting populations. [30, 31] This may be accomplished by modifying the expression $N_{t-1}$ in equation 3 for each clone, by taking the sum of $N_{t-1}$ for all the other competing T cell populations when calculating $N_t$ for each clone. Each clone's $N_{t-1}$ is weighted by a correction factor $\alpha$ for its interaction with the T cell clone being examined. Given the central role for the target mHA-HLA complex's $IC50$ in determining the T cell frequency for each clone, $\alpha$ for each clone is calculated by dividing the $IC50$ of the competing T cell clone with the test clone (note the use of $IC50$ instead of $1/IC50$). This implies that T cell clones recognizing mHA-HLA complexes with a higher binding affinity will have a disproportionately higher impact of on growth of T cell clones binding less avid mHA-HLA complexes and vice versa (Supplementary Figure 3B). The resulting square matrix will have a series of 1s going down the diagonal and values <1 above the diagonal and >1 below that.

|        | $TC_1$          | $TC_2$          | $TC_3$          | $TC_n$          |
|--------|-----------------|-----------------|-----------------|-----------------|
| $TC_1$ | $IC50_1/IC50_1$ | $IC50_1/IC50_2$ | $IC50_1/IC50_3$ | $IC50_1/IC50_n$ |
| $TC_2$ | $IC50_2/IC50_1$ | $IC50_2/IC50_2$ | $IC50_2/IC50_3$ | $IC50_2/IC50_n$ |
| $TC_3$ | $IC50_3/IC50_1$ | $IC50_2/IC50_3$ | $IC50_3/IC50_3$ | $IC50_3/IC50_n$ |
| $TC_n$ | $IC50_n/IC50_1$ | $IC50_n/IC50_3$ | $IC50_n/IC50_3$ | $IC50_n/IC50_n$ |

To account for *n* competing T cell populations, equation 3 will be modified as follows

$$N_{t\,TCx} = \frac{K^{1/IC50mHAx}}{(K^{1/IC50\,mHAx} - \sum_1^n N_{t-1\,TCi} \cdot \alpha_i)(e^{rt(1/IC50mHAx)}) + 1} \qquad [3.1]$$

Where for the T cell clone $x$, the $N_t$ will depend on the sum of the clonal frequencies of all the $n$ T cell clones in the previous iteration, with the $\alpha$ of each T cell clone $i$, with respect to the T cell clone $x$, modifying the effect of the frequency of the $i$th clone on $TCx$.

The direction of the T cell clones will be determined by antigen specificity, i.e. TCR identity and whether their receptors recognize recipient mHA-HLA (are alloreactive) or not (are tolerant), such as pathogen directed T cell clones. As an example, a tolerant, non-autoreactive set of T cell clones (donor vector, $\overline{v_{TCD}}$) may be transformed to a predominantly alloreactive set of T cell clones ($\overline{v_{TCR}}$), after interacting with the mHA-HLA complex, alloreactivity operator encountered in the recipient ($\mathbb{M}_{APO}$). On the other hand, a genetically identical twin transplant may not experience a significant transformation of the T cell vector. The pathogen/commensal antigen operator may have a strong formative influence on the eventual T cell repertoire emerging post transplant (Supplementary Figure 3C).

*Whole Exome Sequencing and Variant Peptide Library Generation*

T cell repertoire simulation was performed using data derived from WES of transplant donors and recipients. To accomplish this previously cryopreserved DNA was obtained from donors and recipients of allogeneic SCT after obtaining permission from Virginia Commonwealth University's institutional review board. Nextera Rapid Capture Expanded Exome Kit was used to extract exomic regions from the DNA samples, which were then multiplexed and run on Illumina HiSeq 2500 sequencing machine to achieve an average sequencing depth of ~90 per sample. 2X100 sequencing reads were then aligned to the human reference genome (hg18) using BWA aligner. [32] Single nucleotide polymorphisms (SNPs) unique to the recipient, and absent in the donor were identified using our previously outlined method (Sampson et al 2014), which utilizes best practices for Genome Analysis ToolKit (GATK) [33] along with in-house software. Patients underwent either 8/8 (n=28) or 7/8 (n=6) HLA-A, B, C and DRB1 matched related (n=7) or unrelated (n=27) SCT. HLA matching had been performed using high resolution typing for the unrelated donor SCT recipients; and intermediate resolution typing for class I, and high resolution typing for class II antigens for related donor recipients. In this retrospective study, HLA class I typing information and clinical outcome information was utilized for the patients.

Exome SNP data was used to determine the resulting variant peptide library using Annovar [34] in each transplant DRP. The HLA-A, B and C binding affinity IC50 (nM) was determined for all these peptides by running the NetMHCpan (version 2.8) software, as previously reported. [35, 36] Parsing the NetMHCPan output, unique peptide-HLA combinations were identified and organized in order of declining mHA-HLA affinity. An IC50 cutoff value of 500 nM was chosen to study the various model parameters. IC50 values higher than this cutoff are assumed to represent extremely weak binding and hence, excluded from the downstream analysis. The T cell repertoire simulations were then run in MATLAB (Mathworks Inc., Natick, MA), utilizing the above model (See Supplementary Mathematical Methods and Program). MS Excel was used to graph and visualize output data from the MatLAB analysis. Effect of T cell vector configuration on survival and relapse was examined using SPSS Statistics soft ware package.

**Results.**

*Whole exome sequencing and T Cell Repertoire Simulation*

WES data from 34 patients were used to generate unique, variant peptide-HLA libraries in the GVH direction for each patient using the NetMHC pan software. The mathematical model presented here simulates the T cell repertoire that may result from such minor histocompatibility antigens, associated with exome wide sequence differences between donors and recipients, accounting for their HLA type. The exome difference data is used to generate variant nine-mer peptides, and the resulting mHA-HLA binding affinity specific to an individual DRP is used to populate a square identity matrix. An average, 2260 peptide-HLA complexes (range: 537 to 4755) were evaluated in these calculations. This recipient antigen matrix operator then *transformed* a hypothetical *donor* T cell vector, $v_{TCD}$, with each T cell clone designated to have $N_0$ = 1, representing a single T cell recognizing each of the mHA-HLA complexes. Substituting Equation 3 for the value 1 in the identity matrix $\mathbb{M}_{APO}$ in equation 2, and setting $r$ at 1.5 and $K$ at $10^9$, a sigmoid growth curve is observed in the T cell clonal populations as the iterative calculations are performed to simulate T cell clonal proliferation to get the final $v_{TCR}$. In this model the steady state TC clonal populations depends on the 1/IC50 of the relevant peptide-HLA complex (Figure 1 A-C). In this system, when starting with a single T cell recognizing each mHA-HLA complex, and proliferating in proportion to the binding affinity, only a small number of T cell clones grow to a high frequency. When T cell clones present at a frequency of >10 at *steady state* (iteration beyond which there is no further growth) are plotted on a log-log plot, a high correlation coefficient is seen suggesting a Power Law distribution in these top clones (Figure 2), analogous to T cell clonal distribution observed in normal donors. In keeping with observation from normal donors there are a large number of low copy number clones (<10). Using this model, the sum of all the T cell clones in an individual over time (i.e. after several iterations) results in a sigmoid curve, analogous to the sigmoid growth behavior in post allograft setting when lymphocyte populations have been examined (Figure 3).

The resulting T cell repertoire is similar to that seen clinically, and demonstrates features such as power law distribution and logistic kinetics over time. When the derivative is determined for the population of various clones it coincides exactly with each clonal curve, with dN/dt rising until the exponential growth curve reaches its midpoint and declining thereafter until it reaches essentially 0 when the growth curve plateaus at K (Supplementary Figure 1), as would be expected. Importantly this model demonstrates that despite a very large array of peptide mHA, only a few T cell clones become dominant; thus identifying peptides that appear to be of greatest relevance under the 'chosen' conditions, but more importantly giving a likeness of normal clonal distribution. This behavior of the model generated T cell repertoire is analogous to the *in vivo* difference between antigen driven and homeostatic T cell proliferation. T cell clones bearing the receptors which recognize target antigens, when stimulated by relevant second signals, will result in an antigen driven exponential expansion to the carrying capacity of the system, while the remaining T cell clones which do not have a relevant antigen or only have a weakly HLA bound antigen will fail to expand, but will survive at low numbers, so-called, homeostatic expansion, under ambient conditions. When scrutinizing these results it is important to recognize that at the current stage of investigation, the source data for this calculation, the IC50s of peptide-HLA complexes does not filter for GVHD target organ protein expression, thus this model has a hypothesis generating significance primarily.

*Competition between T cell clones leads to variability*

The T cell clones do not proliferate in isolation, rather they are doing so in a milieu of other T cells responding to different antigens, each with varying binding affinity to HLA molecules. Accounting for competition between T cell clones introduces significant variability in the clonal populations during the early iterations following growth initiation (Figure 4 A-C), with specific clonal populations growing rapidly and then plummeting only to grow back again with successive iterations. This effect of competition was most evident on clones responding to low affinity peptides, where the clones were effectively suppressed, only to emerge with later iterations. Therefore as the number of iterations increased so did the number of clones with a detectable population, in other words with increasing number of iterations, a larger number of T cell clones populated the curves, consistent with the physiological observation of increasing T cell repertoire complexity, diversity with the passage of time following SCT. Similar to the case of noncompeting clones, the TCR vector configuration again followed a Power Law distribution, and because of the initial transient suppressive effect of competition on lower affinity clones (Figure 5) over time the repertoire size increased. Due to the variability observed in the individual clones, the sum of all clones for each individual demonstrated marked variability, with the dominant clones suppressing weaker clones early on and weaker clones emerging later on (Figure 6). Initial overgrowth was followed by later decline to a steady state value, in individual clones and the overall repertoire which fluctuated about an average value slowly increasing over time as the lower affinity clones started to grow. As a larger number of clones are incorporated there is a significant increase in the variability of the curve with TC clones directed at high affinity peptides demonstrating early variability and lower affinity peptide targeting TC clones experiencing late variability. This chaotic behavior is best illustrated by constructing phase space plots (Figure 7), where successive values of the clonal population are plotted against each other ($N_{t+1}$ plotted against $N_t$). The phase space plots, with the designated parameters ($N_0$, $r$, $K$) demonstrate clonal fluctuation (loops in the plots) each eventually settling down to an *attractor*, a limited region of the phase space that the system tends to evolve towards. Striking similarity is seen in the behavior of the different clones, albeit at different scales of magnitude, supporting the observation of self-similarity of the repertoire across scales of magnification. The self-similar repertoire distribution of these simulations was further supported by Log-Rank analysis of the repertoire at different time intervals. There was general conformity to a Power Law growth patterns, when log-rank analysis was used to analyze the data sets for individual patients, after the 50th, 100th and >2000th iterations (Figure 8). These analyses provide a possible way to compare model output with real-life TC clonal frequency data from high throughput sequencing.

*Vector-Operator modeling reveals different alloreactivity potential in HLA matched individuals*

When WES-derived HLA-peptide binding IC50 data was evaluated using the competing T cell clonal vector-alloreactivity operator model, the HLA matched patients could be differentiated into three different categories, according to the T cell vector configuration. These were termed $\alpha$, $\beta$ and $\gamma$ for reference. The T cell vector configuration $\alpha$ was characterized by a rapid, exponential rise in the sum of all clones to a value >200,000 when the K was set to $10^9$, whereas $\beta$ was characterized by a slower growth to a level between 200,000 and 20,000, and finally the slowest growing, $\gamma$, configuration which remained at <20,000 T cells at steady state (as many iterations as the number of peptide-HLA complexes) (Figure 9). In the simulations run to date, the groups $\alpha$ and $\beta$ had relatively lower variability in the clonal and total T cell populations in the later iterations, while the group $\gamma$ demonstrated the greatest late variability, perhaps reflecting the relatively larger impact of the lower affinity peptide directed T cell clonal growth, in this relatively low alloreactivity potential system. This difference observed in the model output is likely related to the number of mHA and more importantly the distribution of the binding affinity of the variant peptides to the HLA in the unique donor recipient setting. Further, this finding of a variable WES derived estimate of alloreactivity potential in similarly HLA-matched individuals may provide an alternative explanation of the variability observed in the incidence of alloreactivity following SCT.

Evaluation of the demographics of the patients with different vector configurations demonstrates some surprising observations, for example the vector configuration $\gamma$ is seen more commonly in patients with HLA mismatched donors and in patients of African ancestry, and was no less common in gender mismatched, or unrelated transplantation (Table 1). In our cohort of patients in whom exome sequencing was performed retrospectively, a trend towards survival advantage and reduced relapse was discernable in patients who had a vector configuration $\alpha$ and $\beta$ (Figure 10).

*Properties of the T cell vector and alloreactivity operator system*

This system of simulating alloreactive CD8+ cell repertoire using the mHA-HLA complex operator does not predict GVHD (data not shown). This may be related to the limited nature of mHA panel studied (only nona-mers on HLA class I molecules), and the absence of GVHD target organ specific protein expression data to make the alloreactivity operator matrix closer to reality. In order to test the versatility of the matrix system an arbitrary set of multipliers was developed to study the effect of large-scale organ distribution, particularly on the low affinity peptides (see mathematical methods). When these calculations were performed, a significant increase in the magnitude of all the T cell clones was observed, and while Power Law relationships were maintained the number of clones showing a large magnitude of growth was higher. This also slowed down the rate of rise of the clonal populations, and as can be seen (Figure 11) there was a substantial increase in the magnitude of the final sum of all clones, albeit at the expense of greater variability, less rapid growth observed in the different curves.

A very important property of this model is its ability to alter the direction and magnitude of a donor TCR vector to a new recipient vector. In other words clones that are dominant in one instance will be down-regulated and low frequency clones will be up-regulated if they encounter their cognate antigens with a higher binding affinity. In the model this is demonstrated by taking the output of one individual's steady state T cell repertoire, and inverting them such that the K for the least avid peptide-HLA complex directed TC clone becomes the $N_0$ for the most avid complex. The matrix is then iterated till it reaches a steady state. This results in a complete inversion of the clonal frequencies (Figure 12). This finding is consistent with donor TCR repertoire being altered upon exposure to recipient derived antigens following SCT.

**Discussion**

HLA matching is the gold standard for determining histocompatibility in SCT recipients, however in recent years the use of alternative donors has become increasingly common, with changes in post transplant immunosuppression, allowing transplantation across HLA barriers. GVHD risk with alternative donor transplants is similar to, if not somewhat lower than conventional HLA matched transplants. This implies that alloreactivity in the form of GVHD is determined, in large part by sequence differences outside the major histocompatibility locus, in other words minor histocompatibility antigens. There is a large body of exome sequence differences, which contribute an equally massive array of potential peptide antigens in individual transplant DRP. This array of antigenic targets will have a formative influence of the T cell repertoire and overall immune reconstitution in the recipient after SCT, which may in turn influence clinical outcomes. In this paper, dynamical system modeling using the notion of vectors and operators from matrix mathematics demonstrates that different DRP have significant differences in their ability to generate an alloreactive T cell response. This may allow prediction of the likelihood of alloreactivity developing in specific transplant donors and recipients.

Immune reconstitution following SCT is a dynamical process where the components of the system evolve as a logistic function of time. Making a basic assumption that T cell expansion will be governed by the binding affinity of the variant peptide to HLA, a putative donor T cell response to the variant peptide library in the recipient may be simulated. This simple *T cell vector-alloreactivity operator* model reproduces many of the observations from recent studies of T cell repertoire organization. Specifically, assuming T cell clonal growth in proportion to the target antigen-HLA binding affinity, both individual T cell clones and the entire T cell repertoire follow a sigmoid, logistic dynamic over time as the system iterates. This is similar to immune reconstitution kinetics in a cohort of allograft recipients reported recently [4]. While this behavior is inherent to the use of a logistic equation to model the growth of each clone, the generalization to the entire repertoire results from the using the rule that steady state T cell population of specific clones will be proportional to the binding affinity of the variant peptides to HLA. In essence, the more strongly bound a peptide the more likely it is to elicit a strong response, obviously in the presence of appropriate second signals. Further validation of the model simulating normal physiology comes from the resulting steady state T cell repertoire approximating a Power Law distribution of the T cell clonal frequency for the entire repertoire [18, 19, 20]. The model predicts the emergence of a limited number of dominant T cell clones, symbolizing antigen driven proliferation and a large number of low frequency clones representing a reservoir of T cell clones which are likely sustained through homeostatic mechanisms. [37, 38, 39] Recent observation of similar T cell receptor VDJ expressing clones dominating in different patients with a similar HLA type also lends credence to this model and its underlying premise. [40] Incorporating competition, with high affinity antigen directed T cell clones dominating lower affinity antigen directed clones (Supplementary Figure 3B), introduces instability with wide fluctuations in individual T cell clonal frequency which results in variability in the early T cell repertoire, and chaotic behavior. In the model incorporating competition the dominant higher affinity T cell clones suppress the lower affinity T cell clones early on, however these clones emerge at a later time demonstrating instability at first before eventually settling down. This chaotic behavior is inherent to the competition introduced by multiple clones encountering antigen simultaneously and is a property of the Lotka-Volterra modification used to calculate individual clonal growth. [41] Similar behavior has recently been reported where wide variation has been demonstrated in the T cell clonal population over time following transplantation. [42] Clonal repertoire expansion with time is also observed, similar to increasing repertoire complexity observed following SCT. Finally, this model provides a quantitative explanation of the change in T cell clonal distribution changing following SCT. [18, 22, 43]

In patients who have alloreactive antigen driven proliferation dominating from the outset one would then expect an oligoclonal pattern to emerge rather than a more polyclonal repertoire, which may be expected if appropriate normal repertoire reconstitution is occurring. This hypothesis is supported by the earlier observation that patients who have oligoclonal T cell recovery following SCT may be more likely to be susceptible to alloreactivity. [37, 44] Further support for the notion of oligoclonal T cell growth in response to high affinity and abundant minor histocompatibility antigens comes from the observation of clonal growth in mixed lymphocyte reactions (MLR) predicting loss of renal allografts [45]. MLR has also been used to study, alloreactive T cell clonal populations following SCT and has demonstrated the presence of a large number of dominant and low frequency clones over time. [46] It is therefore highly likely that when adjusted for antigen abundance (as demonstrated by the empiric effect of the K multiplier in our model) whole exome sequencing of donors and recipients will be very likely to predict the magnitude of alloreactivity when it is analyzed using a T cell vector-mHA operator model.

The variability and chaotic behavior unveiled when competition between clones is accounted for makes it very likely that the presence of competing, non-cross reactive strongly bound antigens may potentially reduce the likelihood of alloreactivity (Supplementary Figure 3C). This notion is supported by the observation that the TC repertoire diversity is higher in patients without GVHD than in those with GVHD.

Exome Sequencing to Simulate Alloreactive T Cell Reconstitution in SCT; Abdul Razzaq B et al.                    22[40] Competition may be introduced by several different mechanisms, including vaccination, restoration of microbiota and finally epigenetic modification to upregulate cancer specific antigens. Importantly this may allow a non-immunosuppression based mechanism of GVHD prophylaxis. This so because chaotic systems such as the one represented by the T cell reconstitution model with competition are sensitive to initial conditions and the balance of high-affinity alloreactive vs. non-alloreactive clones may be critical in determining likelihood of GVHD developing in these patients.

The complexity of GVHD in SCT, represents a unique model system in which to study immune reconstitution in the background of genetic diversity between the stem cell donors and recipients. Knowledge of the quantitative immunobiology of SCT may then make it easier to develop stem cell donor selection algorithms that improve upon the prevailing standards of donor selection. This dynamical system model of T cell reconstitution following SCT provides a mathematical framework which can be utilized to study the effect of various parameters such as donor T cell dose, immunosuppressive regimens on post transplant outcomes and with further development may lead to more precise titration of T cell dose and immunosuppression to achieve optimal clinical outcomes following SCT, as well as optimal donor selection beyond HLA identity.

**Table 1.** Patient demographics for DRP with different vector configurations (percent).

| | | Vector Pattern (n) | | |
|---|---|---|---|---|
| | | α (10) | β (10) | γ (14) |
| **Diagnosis** | AML/MDS | 8 (80) | 8 (80) | 7 (50) |
| | ALL | 1 (10) | 0 (0) | 2 (14) |
| | CLL | 0 (0) | 1 (10) | 2 (14) |
| | NHL | 1 (10) | 0 (0) | 2 (14) |
| | MM | 0 (0) | 1 (10) | 1 (7) |
| **Donor Type** | MRD | 2 (20) | 3 (30) | 2 (14) |
| | MUD | 8 (80) | 7 (70) | 12 (86) |
| **HLA mismatch** | No | 9 (90) | 10 (100) | 9 (64) |
| | Yes | 1 (10) | 0 (0) | 5 (36) |
| **Patient Age** | <55 | 4 (40) | 7 (70) | 6 (43) |
| | >=55 | 6 (60) | 3 (30) | 8 (57) |
| **Donor Age** | <40 | 6 (60) | 4 (40) | 7 (54) |
| | >=40 | 4 (40) | 6 (60) | 6 (46) |
| **Patient Race** | Caucasian | 9 (90) | 9 (90) | 8 (57) |
| | African American | 1 (10) | 1 (10) | 6 (43) |
| **Race disparity** | No | 9 (90) | 10 (100) | 9 (75) |
| | Yes | 1 (10) | 0 (0) | 3 (25) |
| **Patient/Donor Gender** | Male/Male | 2 (20) | 4 (40) | 7 (50) |
| | Female/Male | 4 (40) | 3 (30) | 3 (21) |
| | Female/Female | 1 (10) | 2 (20) | 3 (21) |
| | Male/Female | 3 (30) | 1 (10) | 1 (7) |

**Figure 1.** Sigmoid growth behavior of simulated individual T cell clonal frequency (non-competing); peptide-HLA binding affinity proportional growth A- IC50 1.8; B- IC50 2.4; C- IC50 3.06. Note declining Y axis values.

A
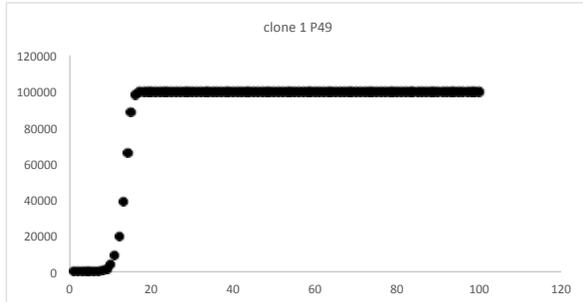

B
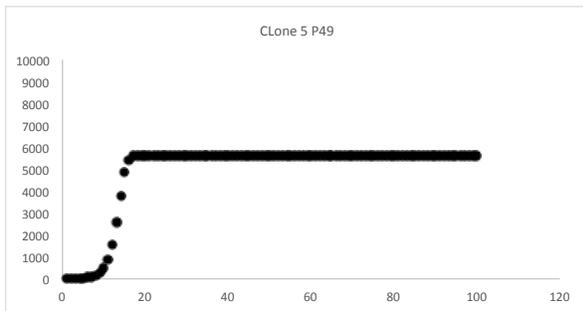

C
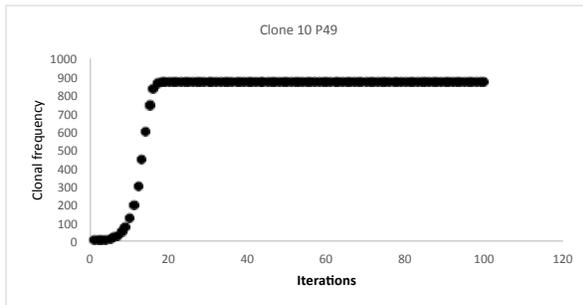

**Figure 2**. Log-Log plot showing Power law distribution of dominant T cell clones (non-competing model). TCR vector transformation after 50 iterations. In this instance a matrix of >2009 T cell clones, each with a $N_0$ of 1 has been 'mapped' to the depicted clonal frequency by the alloreactivity operator. Only T cell clones with a K of >10 are included. Each line indicates a separate iteration of the system. Clonal frequency on the Y axis and individual clones on the X axis.

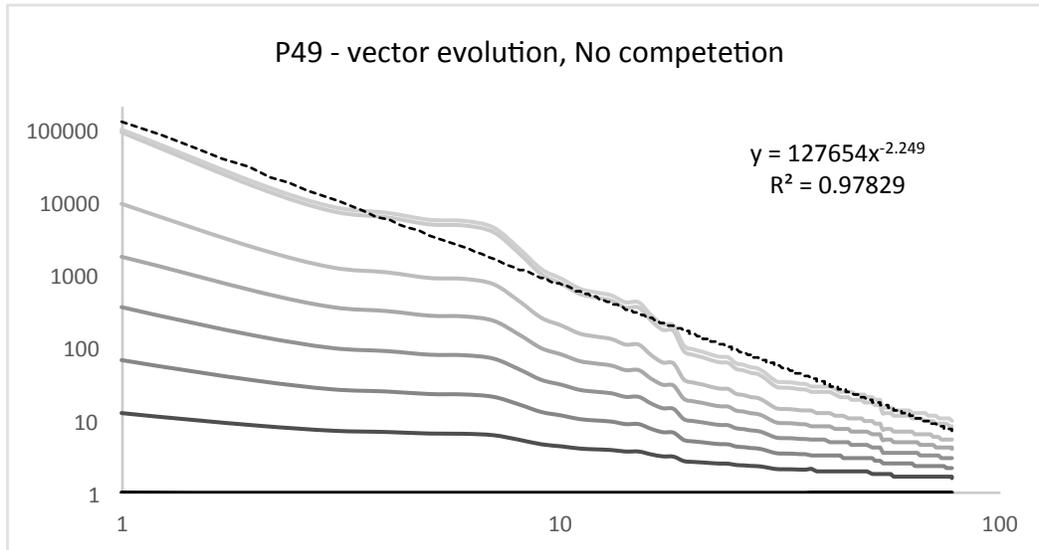

**Figure 3.** Sum of all T cell clones also constitute a sigmoid logistic growth curve, (non-competing growth model)

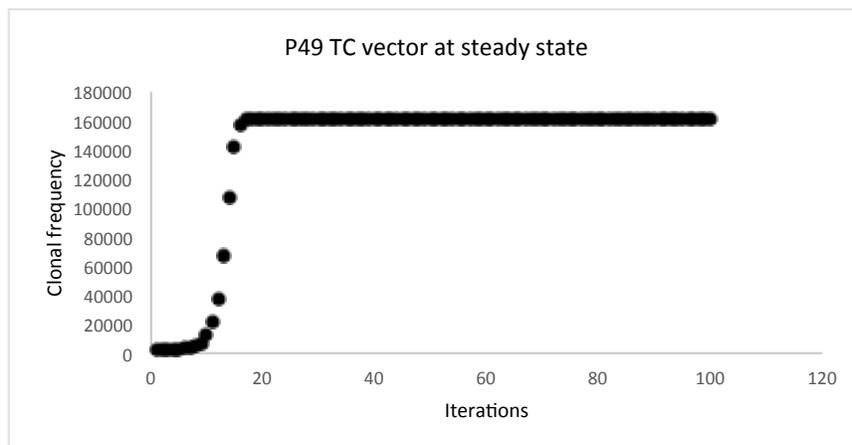

**Figure 4.** Simulated clonal frequency of individual T cell clones accounting for competition between clones. A: IC50 2.49; B: IC50 3.34: C: IC50 4. Marked early fluctuation seen in clonal population, followed by eventual achievement of steady state at a rate proportional to the binding affinity of the target peptide.

A
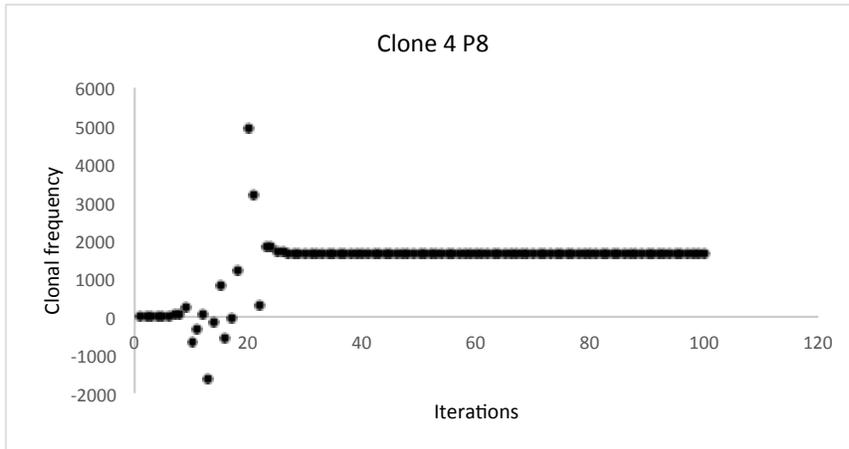

B
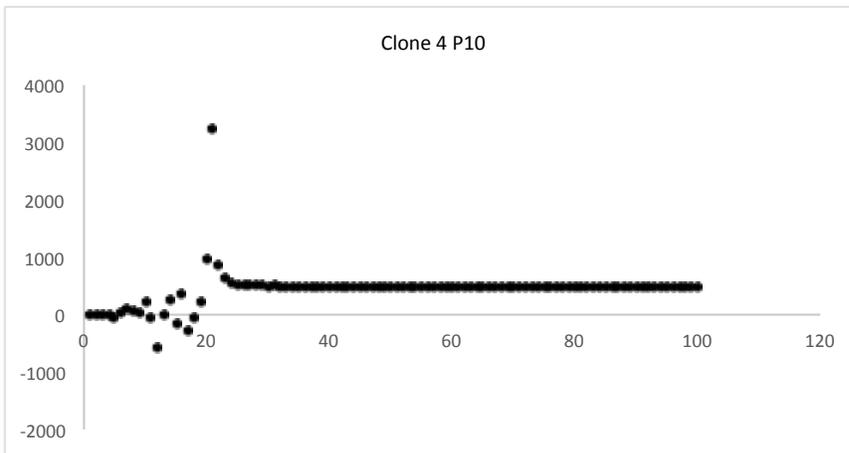

C
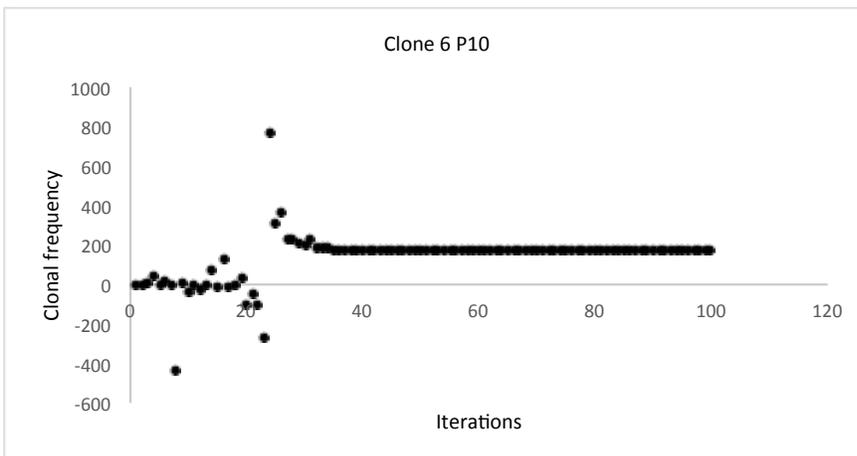

**Figure 5**. Log-Log Plot showing evolution of TC clonal frequency accounting for competition between clones. Data for the T cell clones with a clonal frequency >10 shown for iterations # 25, 50, 75 and 125. Note the progressive increase in clones, with increasing iterations of the system. Number of unique clones on Y axis

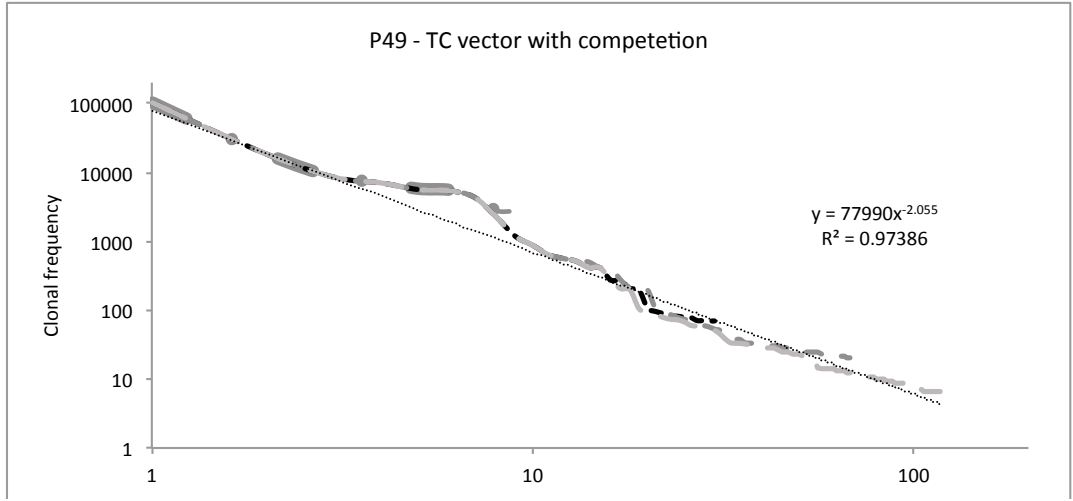

**Figure 6.** Sum of all clones with competition demonstrates the overall sigmoid growth kinetics of the overall TC vector over time, however there is significant variability

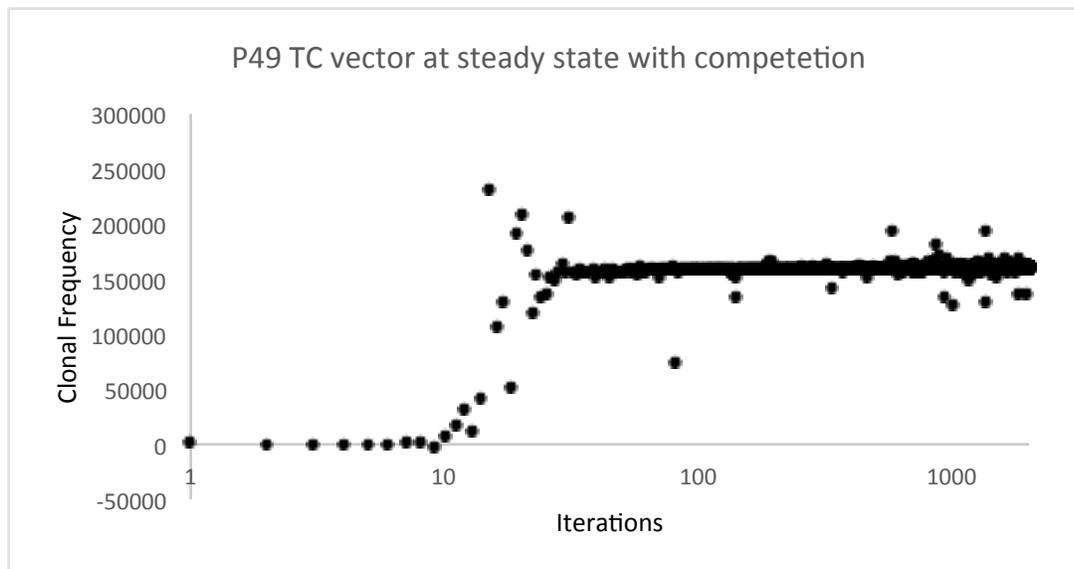

**Figure 7**: Phase space plots of lymphocyte growth patterns, $N_t$ plotted as a function of $N_{t-1}$. Plots A, B and C are successive clones from P33 and D depicts cumulative clones from P47. Clonal frequency loops have a similar morphology in different patients and tend to evolve towards an *attractor* in the phase space.

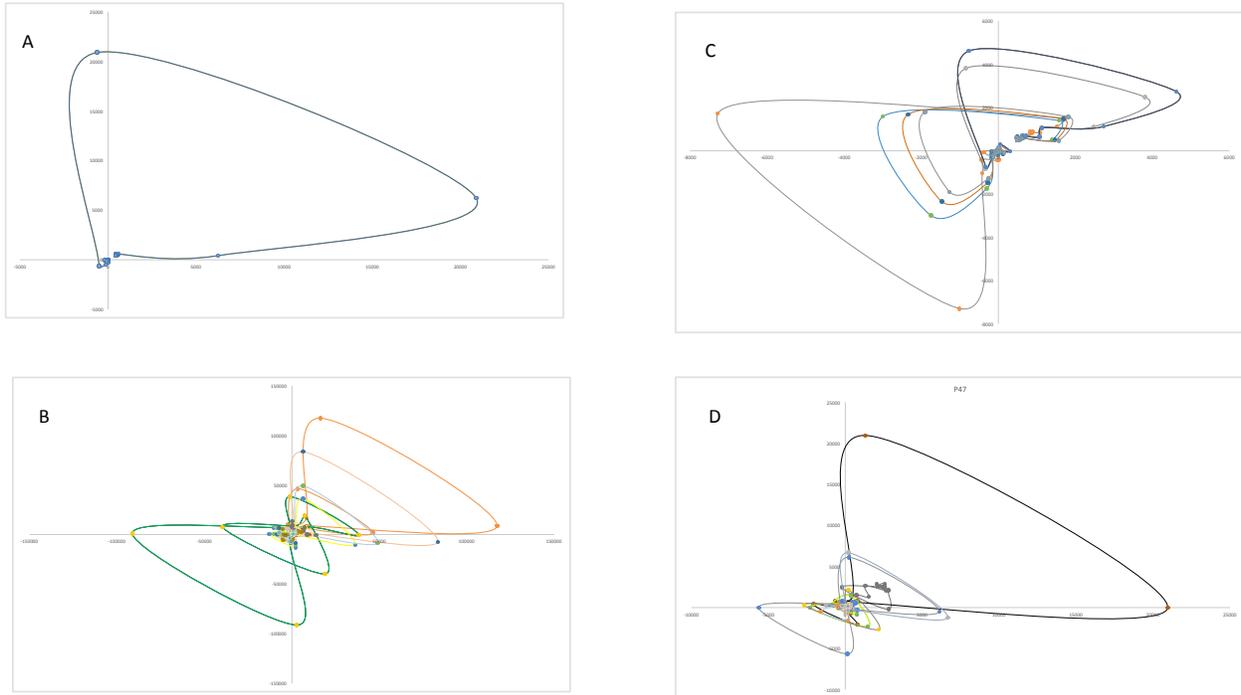

**Figure 8:** Log-Rank Analysis of Simulated T cell clones after 50, 100 and >2000 iterations demonstrates Power Law relationship in the simulations incorporating competetion.

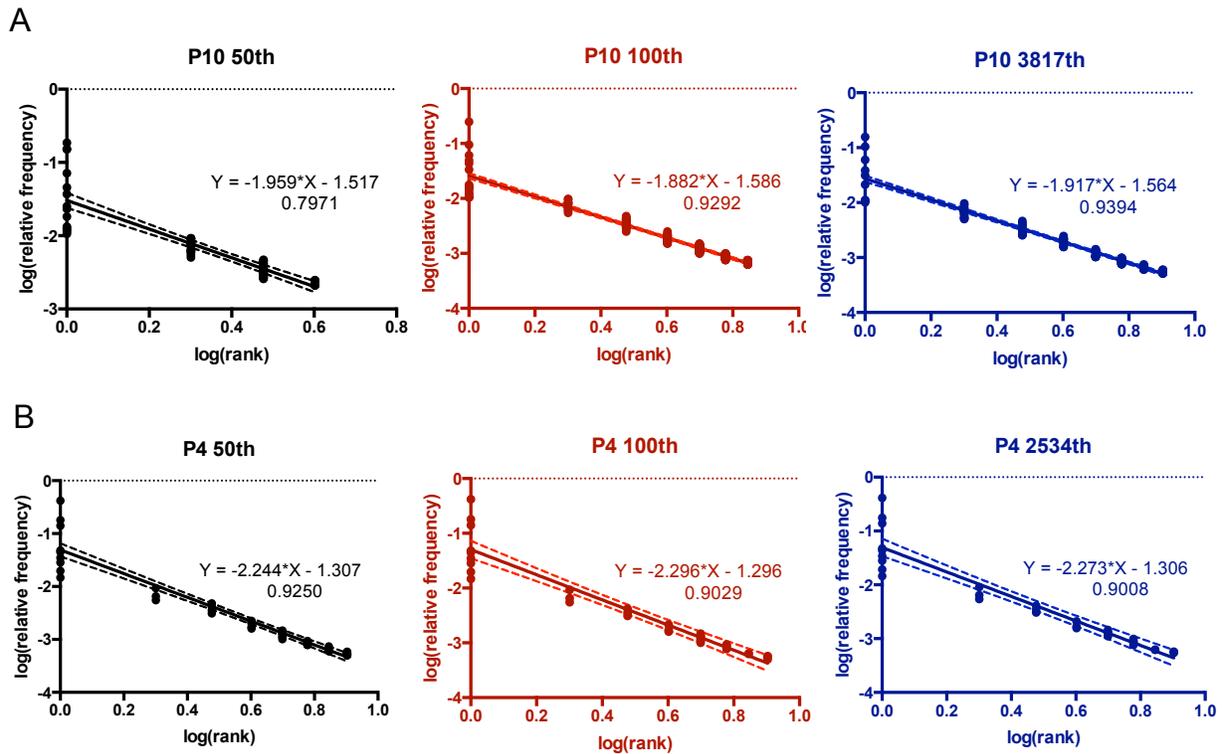

**Figure 9:** Simulated TC repertoire configurations, from different HLA matched patients, fall in three broad classes. α- Rapid average exponential growth with high K; β- Slower exponential growth phase with lower average K; γ- Very slow growth with very low average K value

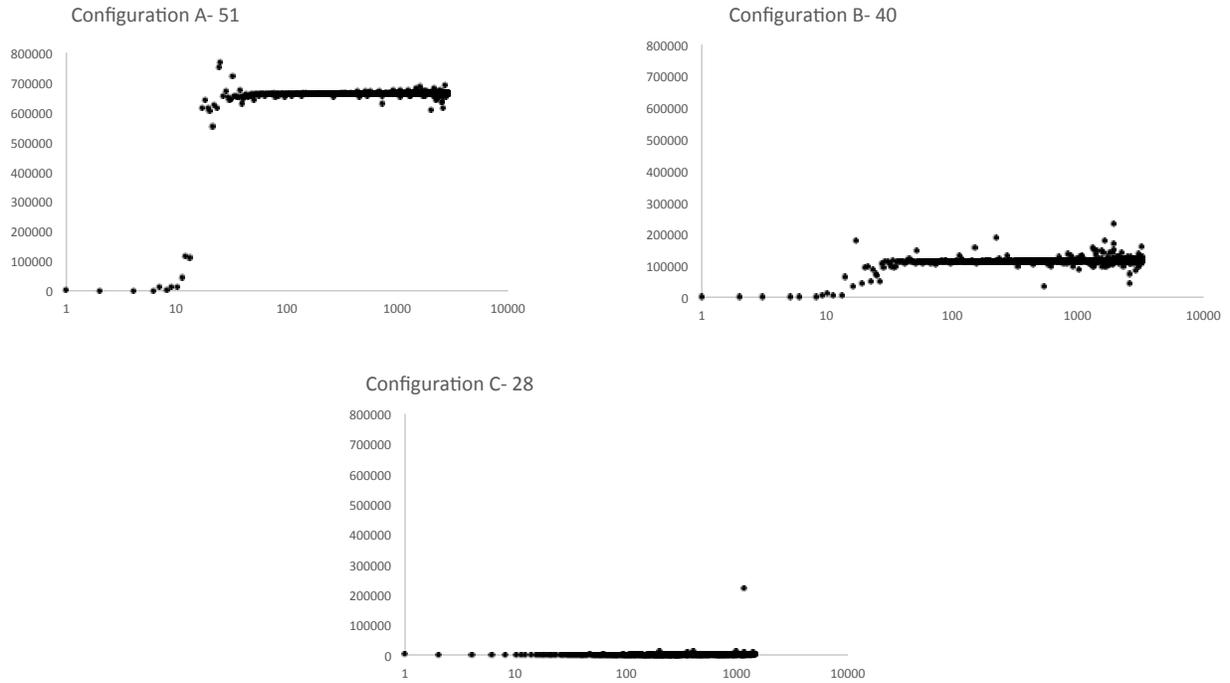

**Figure 10:** Survival (A) and relapse (B) in patients with TCR vector configuration α, β and γ respectively (determination based on the 100<sup>th</sup> iteration). Mantel Cox, test for equality of distribution Log Rank P =0.079 for survival, and P=0.195 for relapse

A

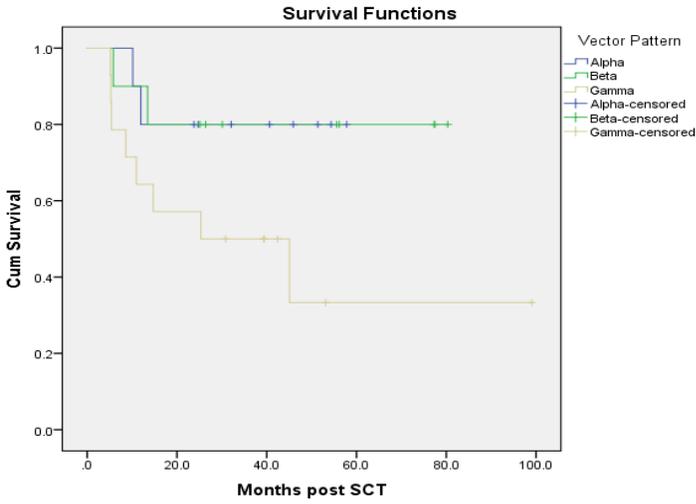

B

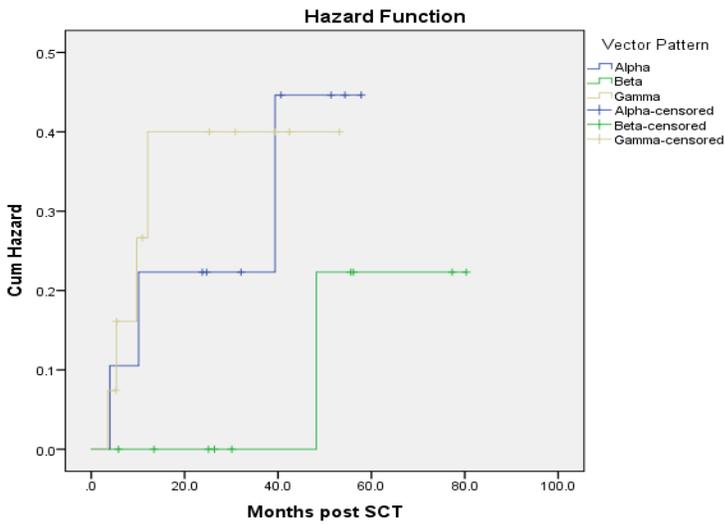

**Figure 11**: Cumulative effect of multipliers on K for individual clones to simulate tissue distribution of peptides. Marked slowing of the rate of growth due to a large number of competing clones, and significantly larger magnitude of eventual vector. Total clonal frequency plotted against iterations.

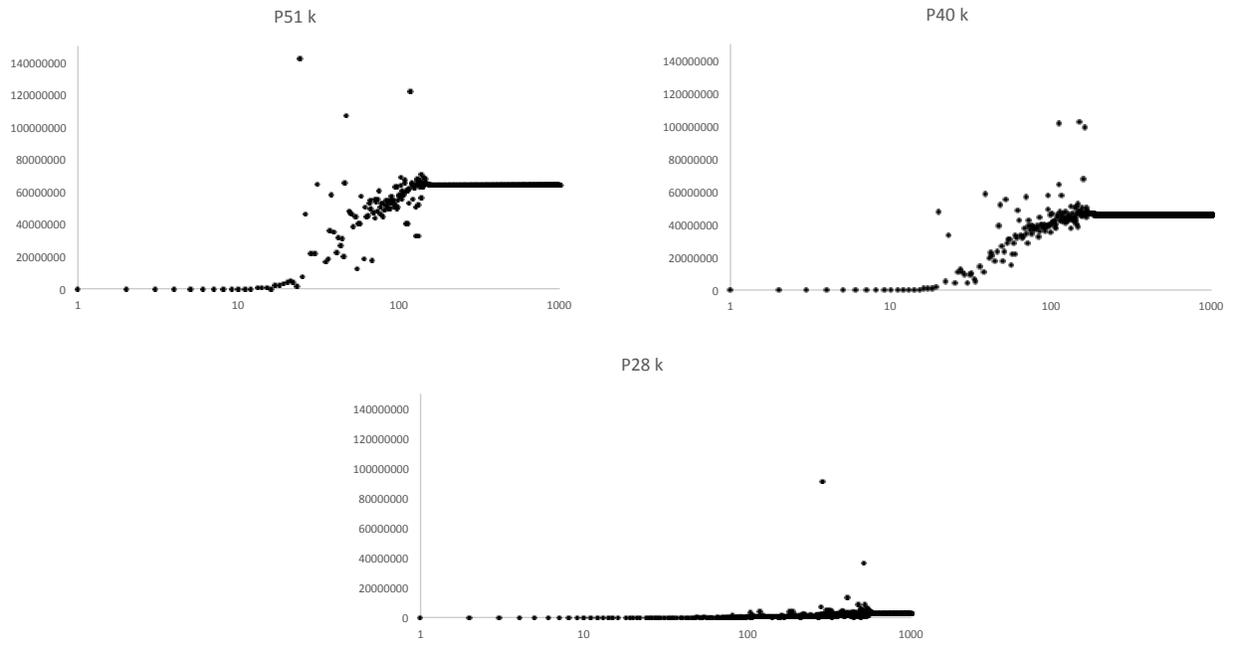

**Figure 12:** Vector transformation by the alloreactivity operator. Original T cell vector (Gray) mapped to a new configuration (black line) when iterated through an inverse of the alloreactivity operator.

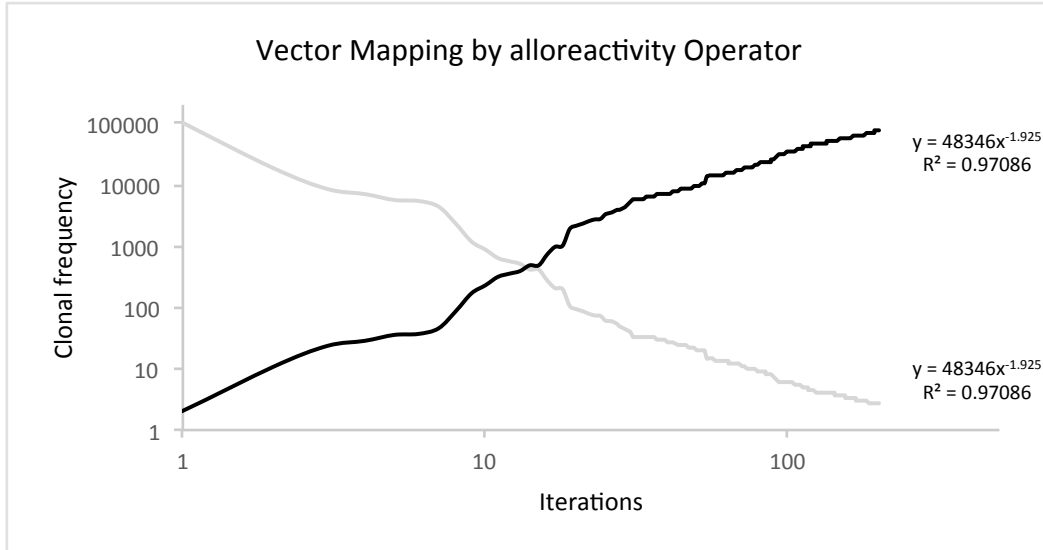

**Supplementary Figure 1:** Logistic expansion, relationship between population size ($N_t$) and instantaneous rate of change of population (dN/dt) at different times during growth. Growth rate rises to a maximum at the midpoint of exponential expansion and then declines until the population reaches K. TC clone responding to peptide with largest 1/IC50 value for patient 2 depicted.

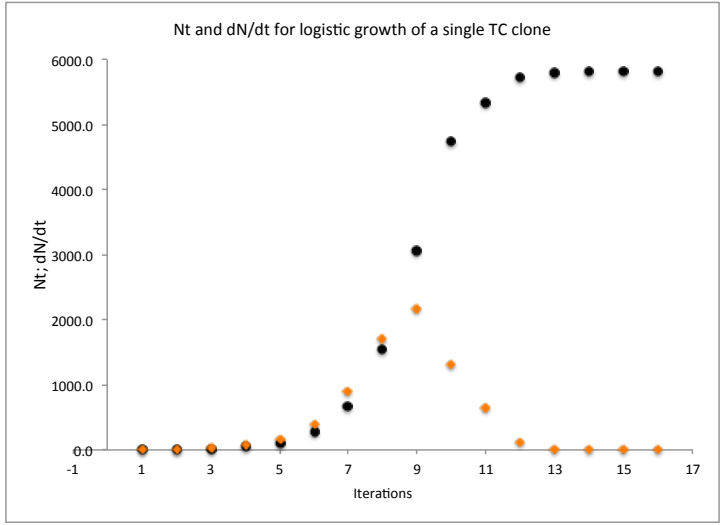

**Supplementary Figure 2:** Vector-operator model of T cell proliferation in response to minor histocompatibility antigen-HLA complexes. T cell clones recognize unique mHA-HLA complexes.

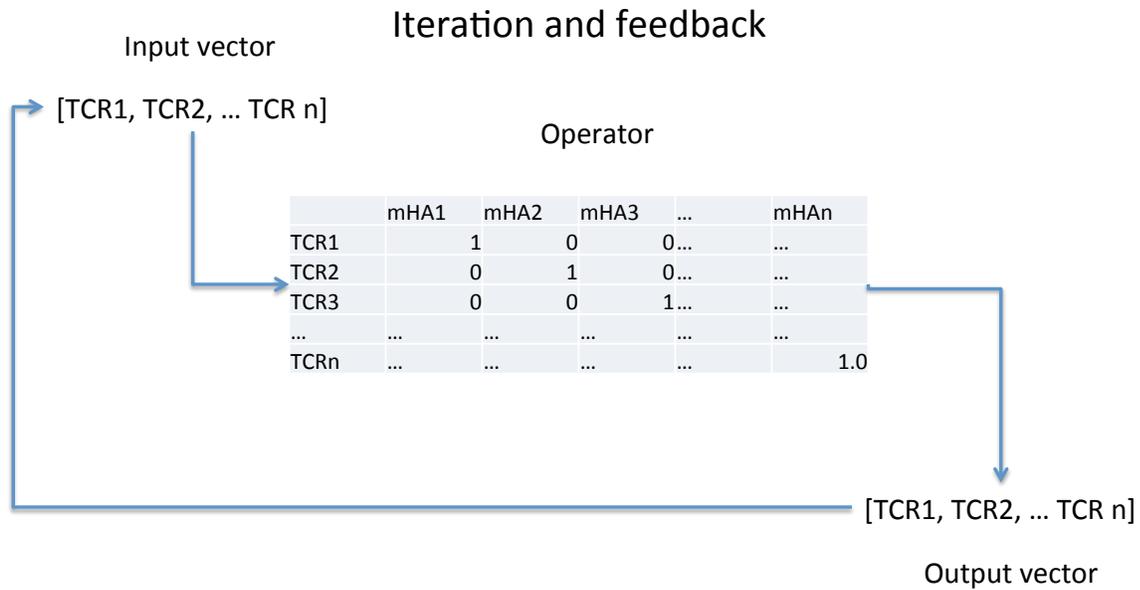

**Supplementary Figure 3:** Graphical representation of logistic growth of T cell clones (closed circles) in response to differing mHA-HLA binding affinity (line atop gray half-frame). Tick marks on X axis represent iterations, and the column of circles represents clonal size at each iteration. A. No competition between clones. B. Competition between clones (width of arrow approximates magnitude of effect of specific clone on others). C. Multiple different sets of antigens contribute to eventual T cell repertoire formation and competition between clones may alter eventual clinical outcomes.

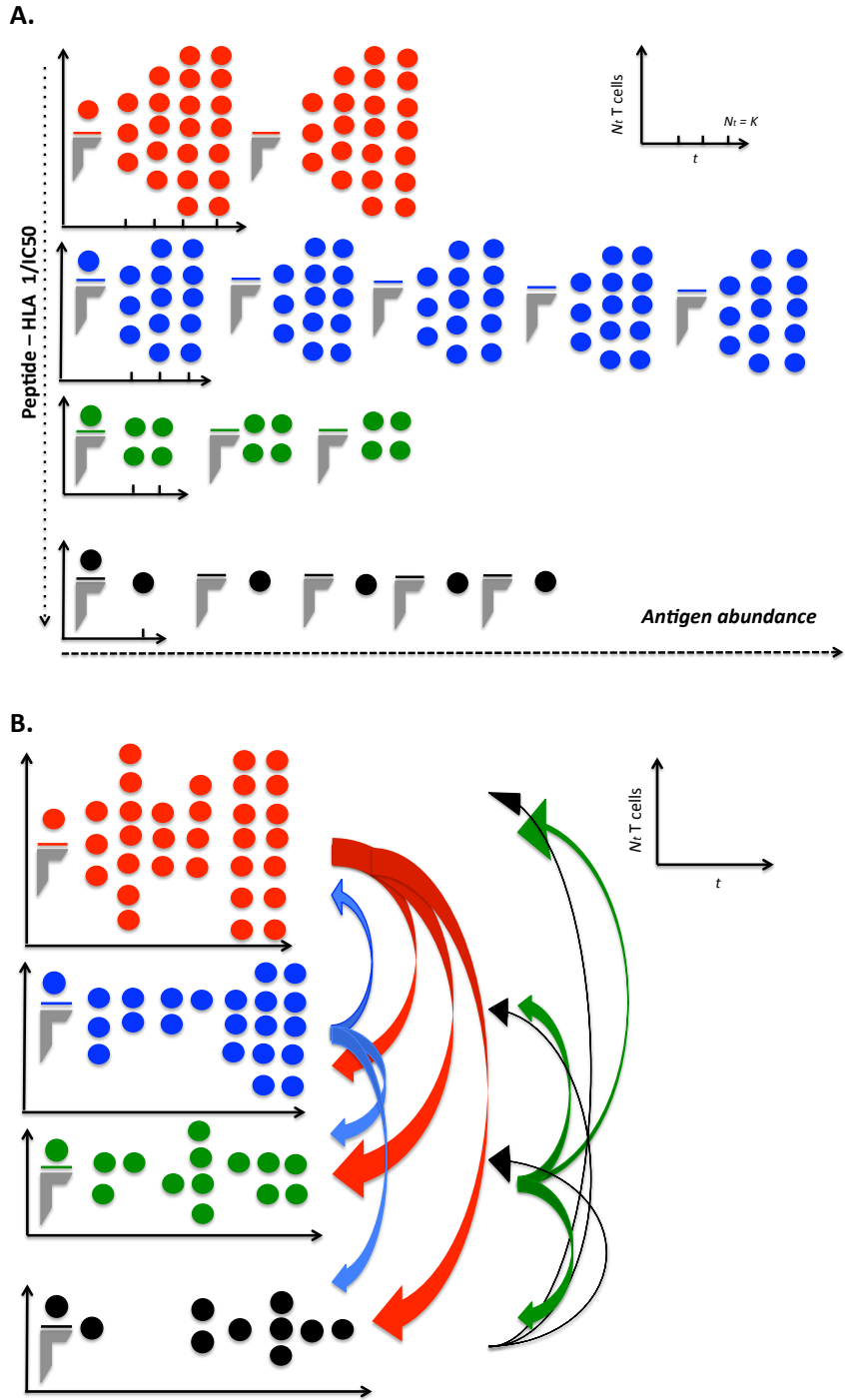

**C.**

Contribution of AP directed T cells (modeled) to whole T cell repertoire – Power Law T cell clonal distribution

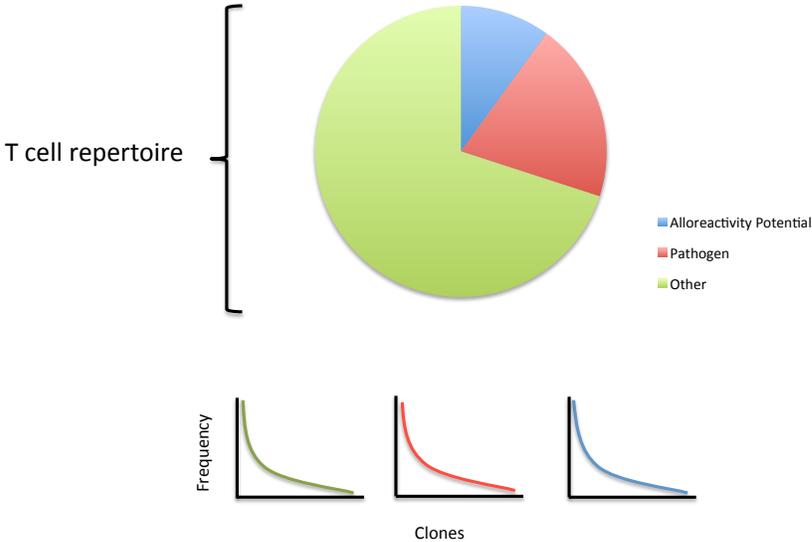

**Supplementary Methods:**

For ease of calculations in the MATLAB and MSExcel the conceptual matrix formulations given in the paper were adapted for use in the named soft ware (Indicated in *italics*).

Model calculations Without Competition.
The non-competing model the uses the vector function as shown in equation below.

$$\overline{v_{TCD}} \cdot \mathbb{M}_{APO} = \overline{v_{TCR}} \qquad [2]$$

Where $\overline{v_{TCD}}$ is the donor vector which is being transformed in the recipient vector $\overline{v_{TCR}}$ using the an operator $\mathbb{M}_{APO}$. *The alloreactivity identity matrix was changed to a single column matrix of IC50 values and incorporated in a single column matrix with the logistic growth formulae incorporated.*
Applying any operation on such a vector this equation maps out the initial and final vectors. The way this complete operation is going to work is when we apply the operator to this vector it transforms $\overline{v_{TCD}}$ to an intermediate form $\overline{v_{INT1}}$ to which the operator is applied again to get $\overline{v_{INT2}}$. This process will be repeated 'n' times till we get a stable vector $\overline{v_{INTn}}$. It should be noted that the vector is a continuously evolving entity and needs to stabilize. The operation has to apply multiple times to allow the vector to be stabilized and that stabilized form $\overline{v_{INTn}}$ will represent the final intended vector which is $\overline{v_{TCR}}$. The following equation to generate the sigmoidal growth of clonal population in a simulated repertoire. Repertoire_nc.m is the program used to develop this model and rerun the vector operation till the vector stabilizes.

$$N_t = \frac{K}{(K - N_{t-1})(e^{rt}) + 1}$$

| IC50 | 1st | | 2nd | | 3rd | | 4th | 5th | | Nth |
|---|---|---|---|---|---|---|---|---|---|---|
| 2.29 | 1 | 1 used as $N_{(t-1)}$ | 3.705158 | 3.705158 used as $N_{(t-1)}$ | 7.132506 | 7.132506 used as $N_{(t-1)}$ | 13.72628 | 26.40666 | …… | 8513.948 |
| 2.53 | 1 | 1 used as $N_{(t-1)}$ | 3.271148 | 3.271148 used as $N_{(t-1)}$ | 5.917546 | 5.917546 used as $N_{(t-1)}$ | 10.69967 | 19.33705 | …… | 3608.38 |
| 2.69 | 1 | 1 used as $N_{(t-1)}$ | 3.047487 | 3.047487 used as $N_{(t-1)}$ | 5.321922 | 5.321922 used as $N_{(t-1)}$ | 9.287676 | 16.1993 | …… | 2216.792 |
| 2.93 | 1 | 1 used as $N_{(t-1)}$ | 2.779801 | 2.779801 used as $N_{(t-1)}$ | 4.637873 | 4.637873 used as $N_{(t-1)}$ | 7.730285 | 12.8755 | …… | 1179.431 |
| 2.93 | 1 | 1 used as $N_{(t-1)}$ | 2.779801 | 2.779801 used as $N_{(t-1)}$ | 4.637873 | 4.637873 used as $N_{(t-1)}$ | 7.730285 | 12.8755 | …… | 1179.431 |
| 4.25 | 1 | 1 used as $N_{(t-1)}$ | 2.009911 | 2.009911 used as $N_{(t-1)}$ | 2.863907 | 2.863907 used as $N_{(t-1)}$ | 4.06477 | 5.761847 | …… | 131.1134 |
| 4.3 | 1 | 1 used as $N_{(t-1)}$ | 1.992843 | 1.992843 used as $N_{(t-1)}$ | 2.828185 | 2.828185 used as $N_{(t-1)}$ | 3.997398 | 5.642726 | …… | 123.8863 |
| 4.38 | 1 | 1 used as $N_{(t-1)}$ | 1.966585 | 1.966585 used as $N_{(t-1)}$ | 2.773561 | 2.773561 used as $N_{(t-1)}$ | 3.894943 | 5.462553 | …… | 113.4474 |
| 4.39 | 1 | 1 used as $N_{(t-1)}$ | 1.96339 | 1.96339 used as $N_{(t-1)}$ | 2.766943 | 2.766943 used as $N_{(t-1)}$ | 3.882575 | 5.440884 | …… | 112.2313 |
| 4.57 | 1 | 1 used as $N_{(t-1)}$ | 1.908937 | 1.908937 used as $N_{(t-1)}$ | 2.65508 | 2.65508 used as $N_{(t-1)}$ | 3.675098 | 5.080035 | …… | 93.18917 |

Where n = number of clones with an IC50 upto 500nm for the respective DRP.

Competition Model:

The competing model uses the same vector operation as the non-competing model but here the vector operator is divided into two components; the alpha matrix and equation 3. The model will operate under similar assumption of the vector continuously changing until it reaches stable values and the operator being applied repeatedly till those stable values are achieved. Following is an explanation of the various components of the operator $\mathcal{M}_{APO}$.

Alpha Matrix

*For ease of performing calculations, the matrix was transposed from the conceptual form described in the methods section of the paper.*

The first step into developing the competing model for TCR is to compute the Alpha Matrix. The alpha matrix is a set of values that act as a weighing factor that determines the clonal strength in a developing repertoire. The simulation computes the alpha matrix for thousands of clones using the $TC_n$ data, but due to obvious limitations an example with just three clones is shown.

$TC_1 = 1$   $TC_2 = 2$   $TC_3 = 3$ (hypothetical values for IC50 of three clones)

Considering $TC_1$ to be the test clone and $TC_2$ to be the competing clone the correction factor would then be

$$\frac{IC50_{TC1}}{IC50_{TC2}} = \frac{1}{2}$$

Hence matrices and established mathematical procedures can be used to calculate the alpha matrix for DRPs in the following manner

$$\begin{bmatrix} IC50_{TC1} \\ IC50_{TC2} \\ IC50_{TC3} \end{bmatrix} \times \begin{bmatrix} \frac{1}{IC50_{TC1}} & \frac{1}{IC50_{TC2}} & \frac{1}{IC50_{TC3}} \end{bmatrix}' = \begin{bmatrix} \frac{TC_1}{TC_1} & \frac{TC_2}{TC_1} & \frac{TC_3}{TC_1} \\ \frac{TC_1}{TC_2} & \frac{TC_2}{TC_2} & \frac{TC_3}{TC_2} \\ \frac{TC_1}{TC_3} & \frac{TC_2}{TC_3} & \frac{TC_3}{TC_3} \end{bmatrix} = \begin{bmatrix} \frac{1}{1} & \frac{2}{1} & \frac{3}{1} \\ \frac{1}{2} & \frac{2}{2} & \frac{3}{2} \\ \frac{1}{3} & \frac{2}{3} & \frac{3}{3} \end{bmatrix} = \begin{bmatrix} 1 & 2 & 3 \\ 0.5 & 1 & 1.5 \\ 0.33 & 0.67 & 1 \end{bmatrix}$$

Where the symbol [ ]' means the transpose of a matrix. The program alpha.m in MATLAB creates these alpha matrices for each DRP prior to running the simulations.

Repertoire:

After the Alpha Matrix has been developed the next step is to generate the repertoire. The repertoire will be developed in a similar fashion but the only difference will be in taking the values of $N_{(t-1)}$.

The model starts by assuming that the repertoire begins with a single cell of each clone and develops from there on. To calculate the second column we will first determine the competing effect of each clone on one another. So we start with the Alpha Matrix of the first ten clones (a 10x10 square matrix). To start with a couple of example. If we were calculating the frequency of $TC_3$ then $TC_1$ has a greater impact on $TC_3$ as compared to $TC_5$ because the affinity is $TC_1 > TC_3 > TC_5$, and as we can see that α for $TC_3$: $TC_1$ is 1.25157 while for $TC_3$: $TC_5$ α has value of 0.85776. N(t-1) is the net effect of the previous clonal population on the currently developing clones hence for a given system the previous population will be multiplied by the corresponding correction factor and the sum of all the corrected populations will be equal to N(t-1).

Considering three clones with IC50 mentioned in the previous section a simulated repertoire is generated below.

| Clone | IC50 | Frequency (t=1) | Frequency (t=2) |
|---|---|---|---|
| $TC_1$ | 1 | 1 | |
| $TC_3$ | 2 | 1 | |
| $TC_3$ | 3 | 1 | |

IC50 for the clones $TC_1 = 1$ $TC_2 = 2$ $TC_3 = 3$. The Alpha matrix for the system of these three clones is

$$\begin{bmatrix} 1 & 2 & 3 \\ 0.5 & 1 & 1.5 \\ 0.33 & 0.67 & 1 \end{bmatrix}$$

Assuming that each clone has one cell at t=1, N (t-1) for each clone will be calculated by multiplying the population by their respective correction factor.
Hence at t=2 to calculate N $_{(t-1)\ (TC1)}$:

$$\left[ [1\ \ 1\ \ 1] \times \begin{bmatrix} 1 \\ 0.5 \\ 0.33 \end{bmatrix} \right]' = [1.88]$$

N $_{(t-1)\ (TC2)}$

$$\left[ [1\ \ 1\ \ 1] \times \begin{bmatrix} 2 \\ 1 \\ 0.67 \end{bmatrix} \right]' = [3.67]$$

N $_{(t-1)\ (TC3)}$

$$\left[ [1\ \ 1\ \ 1] \times \begin{bmatrix} 3 \\ 1.5 \\ 1 \end{bmatrix} \right]' = [5.5]$$

or instead of doing this individually we can do

$$\left[ [1\ \ 1\ \ 1] \times \begin{bmatrix} 1 & 2 & 3 \\ 0.5 & 1 & 1.5 \\ 0.33 & 0.67 & 1 \end{bmatrix} \right]' = \begin{bmatrix} 1.88 \\ 3.67 \\ 5.5 \end{bmatrix}$$

Now using equation 3 we can calculate the simulated TCR.

| Clone | IC50 | Frequency (t=1) | Frequency (t=2) |
|---|---|---|---|
| TC$_1$ | 1 | 1 | 20.08553656 |
| TC$_3$ | 2 | 1 | 4.481573566 |
| TC$_3$ | 3 | 1 | 2.725864415 |

Sample Alpha Matrix

|  | TC$_1$ | TC$_2$ | TC$_3$ | TC$_4$ | TC$_5$ | TC$_6$ | TC$_7$ | TC$_8$ | TC$_9$ | TC$_{10}$ |
|---|---|---|---|---|---|---|---|---|---|---|
| TC$_1$ | 1.00000 | 1.11321 | 1.25157 | 1.45283 | 1.45912 | 1.50943 | 1.52201 | 1.57862 | 1.71698 | 1.93711 |
| TC$_2$ | 0.89831 | 1.00000 | 1.12429 | 1.30508 | 1.31073 | 1.35593 | 1.36723 | 1.41808 | 1.54237 | 1.74011 |
| TC$_3$ | 0.79899 | 0.88945 | 1.00000 | 1.16080 | 1.16583 | 1.20603 | 1.21608 | 1.26131 | 1.37186 | 1.54774 |
| TC$_4$ | 0.68831 | 0.76623 | 0.86147 | 1.00000 | 1.00433 | 1.03896 | 1.04762 | 1.08658 | 1.18182 | 1.33333 |
| TC$_5$ | 0.68534 | 0.76293 | 0.85776 | 0.99569 | 1.00000 | 1.03448 | 1.04310 | 1.08190 | 1.17672 | 1.32759 |
| TC$_6$ | 0.66250 | 0.73750 | 0.82917 | 0.96250 | 0.96667 | 1.00000 | 1.00833 | 1.04583 | 1.13750 | 1.28333 |
| TC$_7$ | 0.65702 | 0.73140 | 0.82231 | 0.95455 | 0.95868 | 0.99174 | 1.00000 | 1.03719 | 1.12810 | 1.27273 |
| TC$_8$ | 0.63347 | 0.70518 | 0.79283 | 0.92032 | 0.92430 | 0.95618 | 0.96414 | 1.00000 | 1.08765 | 1.22709 |
| TC$_9$ | 0.58242 | 0.64835 | 0.72894 | 0.84615 | 0.84982 | 0.87912 | 0.88645 | 0.91941 | 1.00000 | 1.12821 |
| TC$_{10}$ | 0.51623 | 0.57468 | 0.64610 | 0.75000 | 0.75325 | 0.77922 | 0.78571 | 0.81494 | 0.88636 | 1.00000 |

Sample from the generated repertoire.

| IC50 | 1$^{st}$ | 2$^{nd}$ | 3$^{rd}$ | 4$^{th}$ | 5$^{th}$ | 6$^{th}$ | 7$^{th}$ | 8$^{th}$ | 9$^{th}$ | 10$^{th}$ |
|---|---|---|---|---|---|---|---|---|---|---|
| 1.59 | 1 | 6.598178 | 16.94914 | 43.53777 | 111.8365 | 287.2697 | 737.838 | 1894.188 | 4856.994 | 12435.29 |
| 1.77 | 1 | 5.446296 | 12.71195 | 29.67097 | 69.268 | 161.7793 | 378.2893 | 886.4229 | 2090.053 | 5050.477 |
| 1.99 | 1 | 4.516195 | 9.603658 | 20.42694 | 43.49894 | 92.89189 | 199.8581 | 436.7399 | 998.6118 | 2744 |
| 2.31 | 1 | 3.667648 | 7.047683 | 13.56491 | 26.29055 | 51.82635 | 107.0551 | 248.4498 | 993.5055 | -888.138 |
| 2.32 | 1 | 3.647328 | 6.990266 | 13.42023 | 25.95272 | 51.08991 | 105.6466 | 247.2658 | 1060.32 | -780.357 |
| 2.4 | 1 | 3.494855 | 6.565991 | 12.36685 | 23.53661 | 45.96608 | 96.62868 | 249.7243 | 7608.668 | -330.647 |
| 2.42 | 1 | 3.459351 | 6.468938 | 12.13 | 23.00512 | 44.88017 | 94.97819 | 254.6204 | -5716.33 | -275.549 |
| 2.51 | 1 | 3.311107 | 6.071479 | 11.17799 | 20.92262 | 40.8395 | 90.59898 | 321.1369 | -441.993 | -133.192 |
| 2.73 | 1 | 3.014917 | 5.322659 | 9.485205 | 17.56325 | 36.17019 | 116.052 | -228.545 | -63.8262 | -32.5951 |
| 3.08 | 1 | 2.684395 | 4.606455 | 8.147855 | 16.45617 | 59.97837 | -48.6261 | -18.416 | -10.4686 | -5.97809 |
| | 10 | 39.84027 | 82.33822 | 173.9287 | 378.3305 | 872.6914 | 1878.319 | 4291.587 | 11375.54 | 17783.31 |

Organ specific Protein expression.

This model can be easily manipulated to exhibit various properties of a dynamical system. The effect of Organ specific Protein expression can be easily incorporated in this model by introducing a K multiplier. If lower clones come across a large antigen population the proliferation rate of a single cell will not change but the overall effect large populations proliferating in small quantities can make a significant impact on the repertoire.

In the unavailability OSP expression data, such an effect can be simulated using a series of multiplier that will concentrate on increasing the protein expression of the lower clones as opposed to the higher clones.

The multipliers were generated using the first quarter of a sine curve and multiplying the amplitude of that curve with 10000. The figure shows the transformation of a sine curve into a series of multiplier.

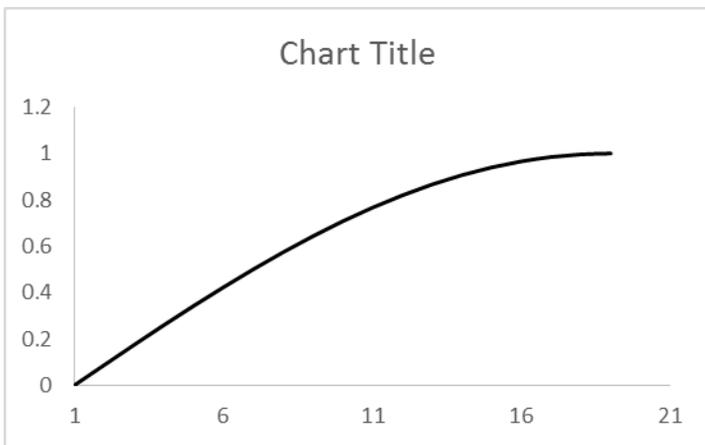
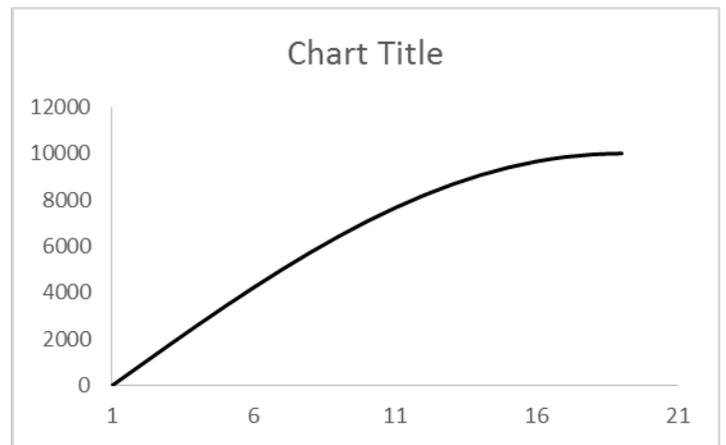

The Clones were then given a serial number and the corresponding multiplier was obtained from this sine curve.
The table below shows the multiplier for the first 10 clones according to the simulated curve shown above

| $TC_1$ | $TC_2$ | $TC_3$ | $TC_4$ | $TC_5$ | $TC_6$ | $TC_7$ | $TC_8$ | $TC_9$ | $TC_{10}$ |
|---|---|---|---|---|---|---|---|---|---|
| 0 | 871.5574 | 1736.482 | 2588.19 | 3420.201 | 4226.183 | 5000 | 5735.764 | 6427.876 | 7071.068 |

The steady state clonal frequency will now be a product of these multipliers and the original steady state values obtained in the previous competing model. Multipliers can be obtained from various mathematical functions. The sine curve was chosen because the impact of a sine curve would more spread out and concentrated towards lower affinity clones as opposed to other mathematical function such as an exponential growth function that will have an impact heavily skewed towards the lower frequency clones.

*Log Rank Analysis*

Clonal rank analysis and subsequent log-log plotting was performed as previously described (Meier et al BBMT 2013). Briefly, the relative clonal frequency from the peptide modeled TCR repertoire was determined by taking the sum of the sequence count for that iteration, and then dividing the sequence count of each unique clone being evaluated by the total sequence count at that level to determine the relative usage of that clone. Rank assignment for clones was based on the prevalence of those clones whose frequency in the overall pool was greater than 0.05%, to determine the organization for the dominant T cell clones in the repertoire. The ranking based on percent frequency (relative frequency × 100, *f*) was done as follows:

$$R_{VDJ} = \begin{cases} 1 & \text{if } f \geq 1\% \\ 2 & \text{if } 0.50\% \leq f < 1\% \\ 3 & \text{if } 0.25\% \leq f < 0.50\% \\ 4 & \text{if } 0.15\% \leq f < 0.25\% \\ 5 & \text{if } 0.10\% \leq f < 0.15\% \\ 6 & \text{if } 0.075\% \leq f < 0.10\% \\ 7 & \text{if } 0.06\% \leq f < 0.075\% \\ 8 & \text{if } 0.05\% \leq f < 0.06\% \\ \text{NA} & \text{otherwise} \end{cases}$$

The difference between the consecutive ranks was consistent with the following progression between ranks: 0.5, 0.25, 0.1, 0.05, 0.025, and 0.01. Log-log plots were created of assigned rank against the relative frequency of those clones, with rank assigned by frequency. A linear relationship between these log-transformed variables is then described by the formula *log y = a log x + log k*, where the slope of the line, '*a*', is considered equivalent to the fractal dimension and is a measure of the developing T cell repertoire.

Program:

Repertoire_NC.m
Reading the IC50 data from the source file.
```
ic = xlsread('a');
```
Creating the affinity data from source data.
```
inv = 1./ic;
```
Initializing rows.
```
x=1;
```
Initializing columns
```
y=1;
```
Initializing timeline 't'.
```
a=1;
c=1;
```
Initializing loop to calculate clonal frequency. The loop condition will equal the number of clones with an IC50 of less then 500nm.
```
while x<200
   b=inv(x,y);
```
Setting the frequency as 1 to initialize the clonal population.
```
   result(x,y)=c;
```
Initializing loop to calculate the clonal frequency of a particular clone throughout the repertoire.
```
   while y<100
```
Adjusting the timeline.
```
   a=a+1;
```
Calculating the steady state clonal frequency.
```
   k=1000000000.^b;
   d=exp(-b*1.5*a);
```
Calculating the clonal frequency using the mathematical model.
```
   hla = (k)/((k-c)*d+1);
```
Saving the current Clonal output as N(t-1) for later use.
```
   c=hla;
   y=y+1;
   result(x,y)=hla;
   end
```
Resetting the values to generate data for following clones.
```
   y=1;
   x=x+1;
   a=1;
   c=1;
end
filename='f.xlsx';
xlswrite(filename,result)
```
The program is set to run row wise. It will generate the clonal frequencies of a $TC_1$ for the whole repertoire and then move on to $TC_2$ and so on.

Alpha_matrix.m
Reading the IC50 data from the source file.
```
ic=xlsread('a');
```
Initializing rows.
```
x=1;
```
Initializing columns.
```
y=1;
```
Initializing columns.
Initializing variables that are used to control row number while calculating Alpha Matrix.
```
r=1;
a=1;
```
Initializing the loop to calculate the Alpha Matrix. Value of y and x represents the total number of clones with an IC50 value less than 500nm
```
wf=zeros(2540,2540);
while y<2541
while x<2541
   e=1;
```
Calculating α
```
   wf(x,y)=ic(r,e)/ic(a,e);
   x=x+1;
   a=a+1;
end
y=y+1;
x=1;
a=1;
r=r+1;
```

```
end
filename=('asa.xlsx');
xlswrite (filename,wf)
Repertoire_c.m
Reading the IC50 data from the source file.
ic = xlsread('a');
Creating the affinity data from source data.
inv = 1./ic;
Initializing rows
x=1;
Initializing columns.
y=1;
Values for the first column of the matrix which represents the beginning populations of the clones.
c=1;
z=0;
Initializing the timeline.
a=1;
'it' represent the one more than the total number of clones.
it=201;
reading the alpha matrix
wf=xlsread('asa');
temporary variable used to calculate the sum of columns.
sum=0;
Initializing matrix to optimize the program
result=zeros(200,1000);
N(t-1) as a matrix calculated for all clones simultaneously
s=zeros(200,1);
Generating the 1st column of the repertoire
while x<it
    result(x,y)=c;
Using temporary variable to calculate the sum of the column.
    sum=sum+result(x,y);
    x=x+1;
end
result(x,y)=sum;
Initializing variable to manipulate alpha matrix
wfa=1;
'y' is the total number of iterations. Can be varied irrespective of the number of clones.
while y<(1000)
Variable that will be used to manipulate alpha matrix and to calculate N $_{(t-1)}$.
    x=1;
    wfa=1;
    d=1;
    sum=0;
    while wfa<it
Loop that runs through the whole column multiplying each cell with its respective factor. This loop runs for single column only where 'wfa' marks the column number
        while x<it
            sum=sum+result(x,y)*wf(x,wfa);
            x=x+1;
        end
        x=1;
Storing the value of N $_{(t-1)}$ to use later.
        s(d,x)=sum;
        sum=0;
The cross product of each column with the clonal frequency gives the value of N $_{(t-1)}$. Although it takes up a lot more time as compared to vector multiplication it provides more flexibility to manipulate the data.
        wfa=wfa+1;
        d=d+1;
    end
Moving through timeline to calculate the clonal frequency from the computed values of N $_{(t-1)}$ stored in the matrix 's'.
    a=a+1;
    y=y+1;
    x=1;
Temporary variable to calculate the sum of the columns.
    sum=0;
    while x<it
        z=y;
        y=1;
        b=inv(x,y);
```

Calculating the current clonal population.
```
    c=s(x,y);
    y=z;
    k=1000000000.^b;
    d=exp(-b*1.5*a);
    hla = (k)/((k-c)*d+1);
    result(x,y)=hla;
    sum=sum+hla;
    x=x+1;

end
result(x,y)=sum;

end
filename='new.xlsx';
xlswrite(filename,result);
```

K_multiplier.m
Reading the IC50 data from the source file.
```
ic = xlsread('a');
```
Creating the affinity data from source data.
```
inv = 1./ic;
```
Initializing rows
```
x=1;
```
Initializing columns.
```
y=1;
```
Values for the first column of the matrix which represents the beginning populations of the clones.
```
c=1;
z=0;
```
Initializing the timeline.
```
a=1;
```
'it' represent the one more than the total number of clones.
```
it=201;
co=1000;
```
reading the alpha matrix
```
wf=xlsread('asa');
```
temporary variable used to calculate the sum of columns.
```
sum=0;
```
Initializing matrix to optimize the program
```
result=zeros((it),co);
```
N(t-1) as a matrix calculated for all clones simultaneously
```
s=zeros((it-1),1);
```
Generating the 1$^{st}$ column of the repertoire
```
while x<it
    result(x,y)=c;
```

Using temporary variable to calculate the sum of the column.
```
    sum=sum+result(x,y);
    x=x+1;
end
result(x,y)=sum;
```
Initializing variable to manipulate alpha matrix
```
wfa=1;
```
'y' is the total number of iterations. Can be varied irrespective of the number of clones.
```
while y<(co)
```
Variable that will be used to manipulate alpha matrix and to calculate N$_{(t-1)}$.
```
    x=1;
    wfa=1;
    d=1;
    sum=0;
    while wfa<it
```
Loop that runs through the whole column multiplying each cell with its respective factor. This loop runs for single column only where 'wfa' marks the column number
```
        while x<it
            sum=sum+result(x,y)*wf(x,wfa);
            x=x+1;
        end
        x=1;
```

Storing the value of $N_{(t-1)}$ to use later.
```
    s(d,x)=sum;
    sum=0;
```
The cross product of each column with the clonal frequency gives the value of $N_{(t-1)}$. Although it takes up a lot more time as compared to vector multiplication it provides more flexibility to manipulate the data.
```
    wfa=wfa+1;
    d=d+1;
  end
```
Moving through timeline to calculate the clonal frequency from the computed values of $N_{(t-1)}$ stored in the matrix 's'.
```
  a=a+1;
  y=y+1;
  x=1;
```
Temporary variable to calculate the sum of the columns.
```
  sum=0;
  while x<it
    z=y;
    y=1;
    b=inv(x,y);
```
Calculating the current clonal population.
```
    c=s(x,y);
    y=z;
```
Each clone is given a rank in a sine curve according to its serial number hence a multiplier of 10000 along with a sine curve ensures that maximum effect is reflected upon the lower clones.
```
    ang = pi/(2*(it-1));
    k=(1000000000.^b);
```
K multiplier used to enhance the clonal frequency.
```
    k=k*(1+(10000*sin(ang*(x-1))));
    d=exp(-b*1.5*a);
    hla = (k)/((k-c)*d+1);
    result(x,y)=hla;
    sum=sum+hla;
    x=x+1;
  end
  result(x,y)=sum;
end
filename='new k2.xlsx';
xlswrite(filename,result);
```

**Acknowledgements.** This work was supported by Massey Pilot Project Grant and an award from Virginia's Commonwealth Health Research Board. BA wrote the program for performing calculations in MATLAB and performed vector-operator calculations presented in this paper along with SS and MM. MS performed sequencing on samples identified and procured by CR and MV. VK performed bioinformatic analysis of the sequencing data to identify unique peptides and their HLA binding affinity. AS created data files with unique peptides and HLA IC50 values and did the statistical analysis as well as collect and verify patient data. JM performed Log Rank analysis. All the authors contributed to writing the manuscript.

Dynamical System Modeling in HLA-Matched SCT    49